\documentclass[aps,prd,onecolumn,showpacs,superscriptaddress,preprintnumbers,floatfix,nofootinbib,reprint]{revtex4-1}

\usepackage{graphicx}
\usepackage{amsmath}
\usepackage{bm}
\usepackage{yhmath}
\usepackage[colorlinks,citecolor=blue,urlcolor=blue,linkcolor=blue]{hyperref}
\usepackage{subfigure}
\usepackage{dcolumn,booktabs,bm}
\usepackage{slashed}
\usepackage{amsfonts,amssymb,stmaryrd,latexsym,amsmath}
\usepackage{txfonts}
\usepackage{multirow}
\usepackage{cancel}
\usepackage{array}

\renewcommand{\arraystretch}{1.8}
\allowdisplaybreaks

\begin{document}

\title{Probing hidden-charm decay properties of $P_c$ states in a molecular scenario}
\author{Guang-Juan Wang}\email{wgj@pku.edu.cn}
 \affiliation{Center of High Energy Physics, Peking University, Beijing 100871, China}
\affiliation{School of Physics and State Key Laboratory of Nuclear
Physics and Technology, Peking University, Beijing 100871, China}

\author{Li-Ye Xiao}\email{lyxiao@ustb.edu.cn}
 \affiliation{University of Science and Technology Beijing, Beijing 100083, China}
\affiliation{Center of High Energy Physics, Peking University, Beijing 100871, China}

\author{Rui Chen}\email{chen$_$rui@pku.edu.cn}
 \affiliation{Center of High Energy Physics, Peking University, Beijing 100871, China}
\affiliation{School of Physics and State Key Laboratory of Nuclear
Physics and Technology, Peking University, Beijing 100871, China}

\author{Xiao-Hai Liu}\email{xiaohai.liu@tju.edu.cn}
 \affiliation{ Center for Joint Quantum Studies and Department of Physics, School of Science, Tianjin University, Tianjin 300350, China}

\author{Xiang Liu}\email{xiangliu@lzu.edu.cn}
 \affiliation{Research Center for Hadron and CSR Physics, Lanzhou University and Institute of Modern Physics of CAS, Lanzhou 730000, China}
 \affiliation{School of Physical Science and Technology, Lanzhou University, Lanzhou 730000, China}

\author{Shi-Lin Zhu}\email{zhusl@pku.edu.cn}
\affiliation{School of Physics and State Key Laboratory of Nuclear
Physics and Technology, Peking University, Beijing 100871,
China}\affiliation{Collaborative Innovation Center of Quantum
Matter, Beijing 100871, China}

\begin{abstract}
  The $P_c(4312)$, $P_c(4440)$, and $P_c(4457)$ observed by the LHCb collaboration are very likely to be $S$-wave $\Sigma_c\bar{D}^{(*)}$ molecular candidates due to their near threshold character. In this work, we study the hidden-charm decay modes of these $P_c$ states, $P_c\to J/\psi p(\eta_cp)$, using a quark interchange model. The decay mechanism  for the $P_c\to J/\psi p(\eta_cp)$ processes arises from the quark-quark interactions, where all parameters are determined by the mass spectra of mesons. We present our results in two scenarios. In scenario I, we perform the dynamical calculations and treat the $P_c$ states as pure $\Sigma_c \bar D^{(*)}$ molecules. In scenario II, after considering the coupled channel effect between different flavor configurations $\Sigma^{(*)}_c\bar D^{(*)}$, we calculate these partial decay widths again. The decay patterns in these two scenarios can help us to explore the molecular assignment and the inner flavor configurations for the $P_c$ states. In particular, the decay widths of $\Gamma(P_c(4312)\to\eta_cp)$ are comparable to the $J/\psi p$ decay widths in both of these two scenarios. Future experiments like LHCb may confirm the existence of the $P_c(4312)$ in the $\eta_cp$ channel.
\end{abstract}
\maketitle

\section{Introduction}

In 2015, the LHCb collaboration reported two pentaquark states $P_{c}(4380)$ and $P_{c}(4450)$ in the $J/\psi p$ invariant mass distribution
of the decay $\Lambda_{b}^{0}\rightarrow J/\psi pK^{-}$ \cite{Aaij:2015tga}. Very recently, with run I and run II data, the LHCb collaboration found that that $P_{c}(4450)^{+}$ should contain two substructures $P_{c}(4440)^+$ and $P_{c}(4457)^+$ \cite{Aaij:2019vzc}. In addition, another new narrow state $P_{c}(4312)^{+}$ is observed. The resonance parameters for these  observed $P_c$ states are \cite{Aaij:2019vzc}
\begin{eqnarray}\label{pcstates}
{P^+_{c}(4312)}: M_{P^+_{c}(4312)}&=&4311.9\pm0.7_{-0.6}^{+6.8}~\text{MeV},\nonumber\\
\Gamma_{P^+_{c}(4312)}&=&9.8\pm2.7^{+3.7}_{-4.5}~\text{MeV},\nonumber\\
{P^+_c(4440)}:M_{P^+_c(4440)}&=&4440.3\pm1.3_{-4.7}^{+4.1}~\text{MeV},\nonumber\\
\Gamma_{P^+_c(4440)}&=&20.6\pm4.9_{-10.1}^{+8.7}~\text{MeV}, \nonumber \\
{P^+_c(4457)}:M_{P^+_c(4457)}&=&4457.3 \pm 1.3^{+0.6}_ {-4.1}~\text{MeV},\,\, \nonumber \\
\Gamma_{P^+_c(4457)}&=&6.4\pm2.0_{-1.9}^{+5.7}~\text{MeV}.
\end{eqnarray}

Before the LHCb's observation, the molecular pentaquark states have been predicted in Refs.~\cite{Yang:2011wz,Wu:2010jy, Wang:2011rga,Wu:2012md,Karliner:2015ina}.~The observation of $P_c$ states by the LHCb collaboration in 2015 inspired theorists' great enthusiasm on the study of hidden-charm pentaquark states. Various interpretations have been proposed, such as the loosely bound meson-baryon molecular states \cite{Chen:2019asm,Liu:2019tjn,He:2019ify,Xiao:2019aya,Meng:2019ilv,Yamaguchi:2019seo,Valderrama:2019chc,Liu:2019zvb,Huang:2019jlf,Wu:2019adv,Sakai:2019qph,Guo:2019kdc,Xiao:2019mst,Chen:2019bip,Voloshin:2019aut,Guo:2019fdo,Lin:2019qiv,Gutsche:2019mkg,Burns:2019iih,Wang:2019hyc,Du:2019pij,Wang:2019ato}, the tightly bound pentaquark states \cite{Weng:2019ynv,Ali:2019npk,Ali:2019clg,Wang:2019got,Giron:2019bcs,Cheng:2019obk,Stancu:2019qga}, and the hadrocharmonium states \cite{Eides:2019tgv}. A recent review is referred to Refs. \cite{Chen:2016qju,Liu:2019zoy,Brambilla:2019esw,Guo:2017jvc,Esposito:2016noz,Brambilla:2019esw,Hosaka:2016pey}. Since $P_c(4312)$ and $P_c(4440)/P_c(4457)$ locate several MeV below the thresholds of the $\Sigma_c\bar D$ and $\Sigma_c \bar D^* $ systems, respectively, the meson-baryon molecule scheme may be a more natural explanation and their parity is proposed to be negative \cite{Chen:2019asm,Liu:2019tjn,He:2019ify,Xiao:2019aya,Meng:2019ilv,Yamaguchi:2019seo,Valderrama:2019chc,Liu:2019zvb,Huang:2019jlf,Wu:2019adv,Sakai:2019qph,Guo:2019kdc,Xiao:2019mst,Chen:2019bip,Voloshin:2019aut,Guo:2019fdo,Lin:2019qiv,Gutsche:2019mkg,Burns:2019iih,Wang:2019hyc,Du:2019pij,Wang:2019ato}. On the other hand, within the models of tightly bound pentaquarks or hadrocharmonium, the masses of three narrow pentaquarks observed by the LHCb are also reproduced and the parity of some states may be opposite  \cite{Ali:2019clg,Wang:2019got,Giron:2019bcs,Cheng:2019obk,Stancu:2019qga,Eides:2019tgv}.  Thus,  more precise data from LHCb are necessary in order to distinguish different hadronic configurations for the $P_c$ states. We should pay more attention to other properties of $P_c$ states, like their decay behaviors, productions, various reactions, and so on.


In particular,  the decay patterns can provide a golden platform for probing their inner dynamics. After the observation of the $P_c(4312)$ and $P_c(4440)/P_c(4457)$, there are several phenomenological investigations on the decay properties of the $P_c$ states as $\Sigma^{(*)}_c\bar D^{(*)}$ molecules by using the heavy quark symmetry \cite{Voloshin:2019aut,Sakai:2019qph}, effective Lagrangian approach \cite{Xiao:2019mst,Lin:2019qiv}, the QCD sum rule \cite{Xu:2019zme} and other methods \cite{Cao:2019kst,Guo:2019fdo}. In this work, we will explore the hidden-charm decay properties of the $P_c$ states in the meson-baryon molecular scheme with the quark interchange model \cite{Wong:2001td,Hilbert:2007hc,Barnes:1991em,Swanson:1992ec,Barnes:1999hs,Barnes:2000hu}. Within the model, the decay widths are related to the scattering process of the hadrons at Born order. The scattering Hamiltonian is then approximated by the well-established quark-quark interactions, with all coupling constants determined by the mass spectra of the hadrons. This method has been adopted to  study  the decay patterns of the exotic states \cite{Wang:2018pwi,Liu:2014eka,Zhou:2019swr}. In this work, we extend the approach to give a quantitative estimate of the decay patterns of the $P_c$ states.

Although these $P_c$ states are probably hidden-charm molecules composed of a charmed baryon and a charmed meson, we have to find out whether they are pure $\Sigma^{(*)}_c\bar{D}^{(*)}$ molecular states or not. In general, the coupled channel effect may play an important role in the systems with the same quantum number and small mass splitting. In the heavy quark limit, the $(\bar D,\bar D^*)$ and $(\Sigma_c,\Sigma^*_c)$ doublets are degenerate, respectively.~There exist strong mixing effects between different flavor configurations $\Sigma^{(*)}_c\bar D^{(*)}$ and they may contribute to the same hidden-charm pentaquark states. In the real world, the mass splittings between different $\Sigma_c^{(*)}\bar{D}^{(*)}$ systems are comparable with the binding energy of the molecular states. They may contribute to the same hidden-charm meson-baryon molecule as the coupled channel effect. For example, in the previous works \cite{Chen:2019asm, Meng:2019ilv,Yamaguchi:2019seo,Xiao:2019aya}, the coupled channel effect is very important to reproduce the three $P_c$ states recently observed by the LHCb collaborations, simultaneously.

In this work, we further explore the influence of the coupled channel effect on the decay behaviors and perform the dynamical calculations on the partial decay widths for the hidden-charm molecular states both as the pure $ \Sigma^{(*)}_c\bar D^{(*)}$ molecules and the admixtures of different flavor configurations.

This paper is organized as follows. After the Introduction, we introduce the formalisms which relate the decay width of the pentaquark state to the effective potentials between the $ \Sigma^{(*)}_c\bar D^{(*)}$ and the $J/\psi p(\eta_c p) $ channels in Sec.~\ref{sec1}. In Sec.~\ref{sec2}, we derive these effective potentials using the quark interchange model. In Secs. \ref{sec3} and \ref{sec4}, we perform the numerical calculations in two scenarios, respectively. In scenario I, the $P_c$ states are treated as pure $\Sigma^{(*)}_c\bar D^{(*)} $ molecules, while in scenario II, they are the admixtures of different flavor configurations. The paper ends with a summary in Sec.~\ref{sec5}. The details of the calculations are illustrated in the Appendices.

\section{Decay width}\label{sec1}

For a decay process of a pentaquark, which we assumed at the moment to be a spinless particle, into a two-body final state $P_c \rightarrow C + D$, its decay width reads
\begin{eqnarray}
d\Gamma=\frac{|\mathbf{p}_{c}|}{32\pi^{2}}\frac{1}{M^{2}}d\Omega_{\mathbf{P}_{c}}|\mathcal{M}_{P_{c}\rightarrow C+D}|^{2},
\end{eqnarray}
where $M$ is the mass of the pentaquark state and $\mathbf{p}_{c}$
is the three momentum of the meson $C$ in the final state. The decay amplitude $\mathcal{M}$ is related to the $T$-matrix as follows \cite{std}:
\begin{eqnarray}
\label{vef}
&&T=\langle\psi_{CD}(\mathbf{p}_{c})|V_\text{eff}(\mathbf{p}_{c},\mathbf{p})|\psi_{P_{c}}(\mathbf{p})\rangle\nonumber\\
&&=-\frac{\mathcal{M}_{p_{c}\rightarrow C+D}}{(2\pi)^{3/2}\sqrt{2M}\sqrt{2E_{C}}\sqrt{2E_{D}}},
\end{eqnarray}
where $E_{C,D}$ is the energy of hadrons  in the final state. The $\psi_{CD}$ is the relative wave function in the final state. In the molecular model, $\psi_{P_c}(\mathbf{p})$ is the normalized relative wave function between the constituent meson $A$ and baryon $B$ in the $P_c$ state. $V_\text{eff}(\mathbf{p_c},\mathbf{p})$ is the effective potential between the $AB$ and $CD$ channels, which is generally a function of the initial  momentum $\mathbf p$ and the final momentum $\mathbf p_c$. At Bonn order, it is derived by the amplitude of the two-body scattering process,
\begin{eqnarray}
A(12)+B(345)\rightarrow C(13)+D(245),
\end{eqnarray}
where the numbers $1-5$ stand for the inner quarks. The detailed derivation is illustrated in Sec.~\ref{sec2}.

In general, a pentaquark may be the superposition of the components with different
orbital angular momentum, and the relative molecular wave function in the momentum space can be expressed as
\begin{eqnarray} \label{rel}
\psi_{P_c}(\mathbf{p})=\sum_{l}R_{l}(p)Y_{lm}(\mathbf{ \hat p}).
\end{eqnarray}

The $T$-matrix is decomposed as
\begin{eqnarray}
&&T={\color{red}}\frac{1}{(2\pi)^3}\int d\mathbf{{p}}\int d\mathbf{k}\delta^{3}(\mathbf{k}-\mathbf{p}_{c})V_\text{eff}(\mathbf{k},\mathbf{p})\psi_{P_{c}}(\mathbf{{p}})\nonumber \\
&&=\frac{1}{(2\pi)^2}\sum_{l}T_{ll}Y_{lm}(\mathbf{ \hat p}_{c}),
\end{eqnarray}
with
\begin{eqnarray}\label{Tll}
&&T_{ll}=\int p^{2}dp\int d u P_{l}( u)V_\text{eff}(p_c,p,u)R_l(p),
\end{eqnarray}
where $u$=cos$\theta$, and $\theta$ is the angle between the $\mathbf{p}$ and $\mathbf{p}_{c}$. $P_l$ is the Legendre polynomial. In this work, we do not include the spin-orbital and tensor interactions.~Then, the orbital angular momentum is kept unchanged in the decay process. $T$-matrix is diagonal in $l$. Then, the decay width is
\begin{eqnarray}
\Gamma & =& \left|\mathbf{p}_c\right|\frac{E_{C}E_{D}}{(2\pi)^3M}\sum_{l}|T_{ll}|^{2}.
\end{eqnarray}
 In the above, we present the equations without considering the spin to illustrate the partial wave analysis of the decay amplitude. For the pentaquark states with spin, we have used a natural generalization of the formulas. The wave function of the pentaquark state reads, 
\begin{eqnarray}
\Psi^{J_{P_c}M_{J_{P_c}}}_{P_c}=\sum_{SM_S,lm}C^{J_{P_c}M_{J_{P_c}}}_{SM_S,lm}[\psi_A\psi_{ B}]^S_{M_S}R_{Sl}( p)Y_{lm}(\mathbf {\hat p}),
\end{eqnarray}
where $J_{P_c}$ ($M_{J_{P_c}}$), $S$ ($M_S$), $l$ ($m$) are the total angular momentum, the total spin and the orbital angular momentum (the third direction component) of the pentaquark state, respectively. $C^{J_{P_c}M_{J_{P_c}}}_{SM_S,lm}$ is the Clebsch-Gordan coefficient. $\psi_A$ and $\psi_{B}$ are the wave functions of the constituent heavy hadrons with  the total spin as $S$.  The radial relative molecular wave function $R_{Sl}$ has two indices since it also depends on the total spin. The decay width reads 
\begin{eqnarray}
\Gamma & =& \left|\mathbf{p}_c\right|\frac{E_{C}E_{D}}{(2\pi)^3M}\sum_{S,l}|T^S_{ll}|^{2}.
\end{eqnarray}
where partial wave amplitude $T^S_{ll}$ is the same as Eq. (\ref{Tll}) once the $R_l$ is replaced by the $R_{Sl}$, which depends both on the spin and the orbital angular momentum.

\section{Effective potential} \label{sec2}
To derive the decay width, we first derive the effective hadron-hadron potential $V_{\text{eff}}$ using the quark interchange model. It is related to the $T$-matrix of the hadron-hadron scattering process $A+B\rightarrow C+D$ at Born order. In the quark interchange model, the scattering process is approximated by the interaction between the inner quarks \cite{Barnes:1991em,Swanson:1992ec,Barnes:1999hs,Barnes:2000hu,Hilbert:2007hc}. In the process from two heavy hadrons scattering into a heavy quarkonium plus a nucleon, the  short-range interactions dominant the scattering process. We adopt the $V_{ij}$ for the quark-quark interaction \cite{Wang:2019rdo,Wong:2001td}. The effective potential $V_{ij}$ in the momentum space can be expressed as,
\begin{eqnarray}\label{quarkmodel}
&&V_{ij}(q^2)= \frac{\mathbf{\lambda}_{i}}{2}\frac{\mathbf{\lambda}_{j}}{2} \left(\frac{4\pi \alpha_{s}}{q^2}+\frac{6\pi b}{q^4}-\frac{8\pi\alpha_{s}}{3m_{i}m_{j}}\mathbf{S}_{i}\cdot\mathbf{S}_{j}e^{-q^2/4\tau^{2}}\right), \nonumber\\
&&\alpha_s(Q^2)=\frac{12 \pi}{\left(33-2 n_{f}\right) \ln \left(A+Q^{2} / B^{2}\right)},
\end{eqnarray}
where $\lambda_{i}(-\mathbf{\lambda}_{i}^{T})$ is the color factor for the
quark (antiquark). $q$ is the transferred momentum. The three terms in the  $V_{ij}$  correspond to the Coulomb, linear confinement, and hyperfine potentials, respectively.~$\alpha_s$ is the running coupling constant and a function of the $Q^2$, which is the square of the invariant mass of the interacting quarks. $\mathbf{S}_{i}$ is the spin operator of the interacting quark. We perform our calculation in the momentum space. The $V_{ij}$ contains a constant potential in the spatial space. In addition, the Fourier transform of the Coulomb and linear confinement potentials induce the divergent terms. The constant term and the divergence vanish due to the exact cancellation of the color factors as illustrated in the following. The parameters in Eq.~(\ref{quarkmodel}) are determined by fitting the mass spectra of the mesons. Their values are listed in Table \ref{qmpara}.

\renewcommand\tabcolsep{0.05cm}
\renewcommand{\arraystretch}{1.8}
\begin{table}[!htbp]
\caption{Parameters in the quark model \cite{Wang:2019rdo,Wong:2001td}.}\label{qmpara}
\begin{tabular}{ccccccccccc}
\toprule[1pt]
& $m_q$ {[}GeV{]}& $m_{c}${[}GeV{]} &
$m_{b}${[}GeV{]} & $b[\text{GeV}^{2}]$ & $\tau${[}GeV{]}  & $A$ & $B${[}GeV{]} & \tabularnewline
&0.334& $1.776$ & $5.102$ & $0.18$ & $0.897$ & $10$ & $0.31$ & \tabularnewline
\bottomrule[1pt]
\end{tabular}
\end{table}

In the quark model, the wave function of a hadron is factorized as
\begin{eqnarray}
\psi=\chi_{c}\chi_{f}\chi_{s}\phi(\mathbf p),
\end{eqnarray}
where $\chi_{c,f,s}$ and $\phi(\mathbf p)$ are the wave functions in the color, flavor, spin and momentum space, respectively. Correspondingly, the $T$-matrix for the scattering process can be factorized as
\begin{eqnarray}
t_{AB\rightarrow CD}
 & \tilde{=}&\mathcal{I}_{\text{color}}\mathcal{I}_{\text{flavor-spin}}\mathcal{I}_{\text{space}},
\end{eqnarray}
where the factors $\mathcal{I}$ with the subscripts $\text{color}$, $\text{flavor-spin}$, and space stand for the overlap of the wave functions in the corresponding space.~The notation $\tilde{=}$ means that the spin and spatial factors can be separated for the $S$-wave scattering process.

\begin{figure*}[!htbp]
\centering
\includegraphics[width=1.0\textwidth]{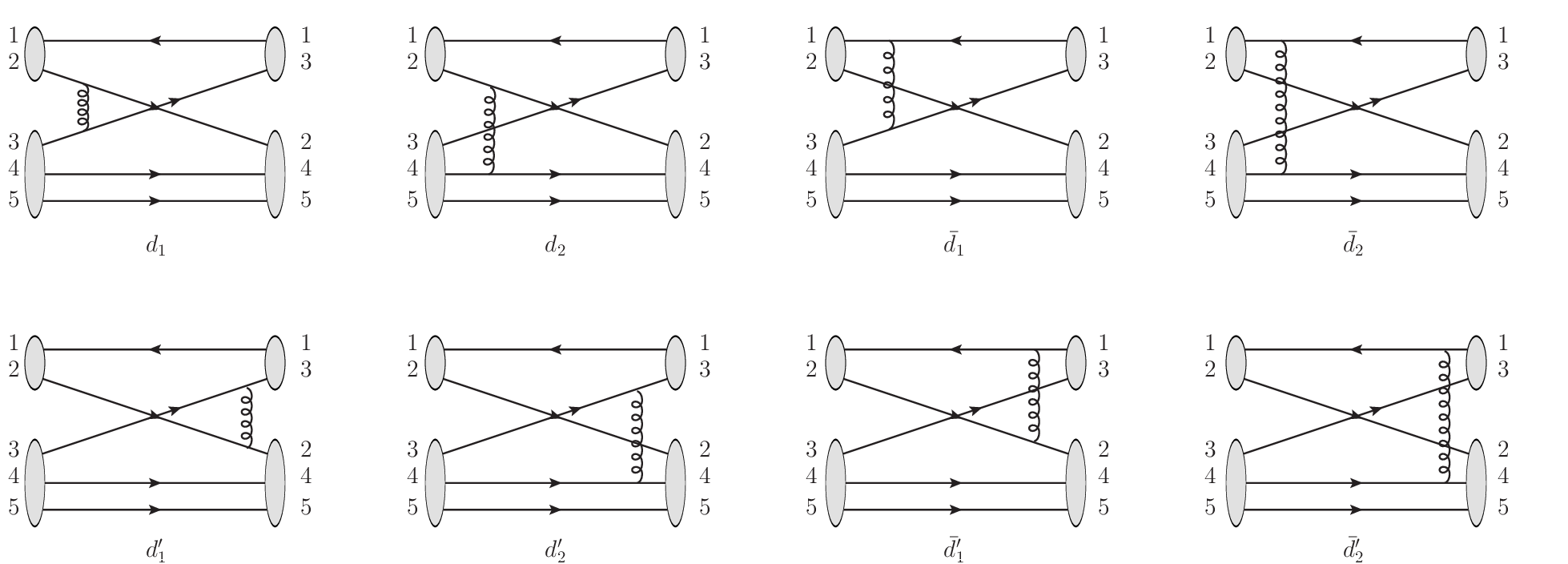}
\caption{ Diagrams for the scattering  process  $\Sigma^{(*)++}_c D^{(*)-}\rightarrow J/\psi (\eta_c) p$ at the quark level. The $d_{i}$ $(\bar d_{i})$ and $d'_{i}$ $(\bar d'_{i})$ diagrams represent the prior and post diagrams, respectively. The curved line denotes the interactions between the quarks. In the nucleon, we use $4$ and $5$ to denote the quarks with the same flavor.  $d_1$ ($\bar d_2$) and $d'_1$($\bar d'_2$) are the same.}
\label{fig:qfill}
\end{figure*}

The so-called  prior-post ambiguity in the scattering process arises due to different decompositions of the Hamiltonian \cite{Swanson:1992ec}. The Hamiltonian is separated as
\begin{eqnarray}
 H=H^0_{A}+H^0_{B}+V_{AB}=H^0_{C}+H^0_{D}+V_{CD},
  \end{eqnarray}
where $H^0$ is the Hamiltonian for a free hadron and $V_{AB(CD)}$ is the residual potential between two color-singlet hadrons. These two decomposition methods result in  the ``prior" and ``post" formalisms as illustrated in Fig.~\ref{fig:qfill}. In the quark interchange model, the $V_{AB(CD)}$ is approximated as the sum of the two-body interactions between the constituent quarks in hadrons $AB$ ($CD$). In the baryon, we mark the light quarks with definite symmetry  as the fourth and fifth quarks. Then, the potential $V_{AB(CD)}$ leads to the four diagrams $(d_1, d_2, \bar d_1, \bar d_2)$ and $(d'_1, d'_2, \bar d'_1, \bar d'_2)$ in the prior and post formalisms, respectively. The prior-post ambiguity disappears if the exact solutions of the wave functions are adopted and the hadrons are on shell\footnote{ For a loosely bound molecule, the two constituent hadrons can be treated as nearly on shell.} \cite{Swanson:1992ec,Schiff}. In this work, we adopt the averaged scattering amplitudes to reduce the prior-post ambiguity.

With the quark interchange model, we first study the scattering process,
\begin{eqnarray*}
&&\Sigma_{c}^{(*)++}D^{(*)-}\rightarrow J/\psi(\eta_{c})p,
\end{eqnarray*}
where $p$ denotes the proton. Then, the scripts $1$, $2$, $3$, $4$, and $5$ in Fig.\ref{fig:qfill} denote the $\bar c$, $d$, $c$, $u$, and $u$ quarks, respectively. The fourth and fifth quarks have the same flavor. Their spin and isospin are equal to one constrained by the Fermi statistics, which simplifies the calculation of the spin-flavor factor $\mathcal{I}_{\text{flavor-spin}}$.

In the quark interchange process, the color factor reads
\begin{eqnarray*}
&&\mathcal I_{\text{color}}=\langle\chi_c^{C}(13)\chi_c^{D}(245)|\frac{\mathbf{\lambda}_{i}}{2}\cdot \frac{\lambda_{j}}{2}|\chi_c^{A}(12)\chi_c^{B}(345)\rangle.\quad
\end{eqnarray*}
In Table \ref{tab:Icolor}, we collect the numerical results of the color factor $\mathcal{I}_{\text{color}}$. Since the sum of the color factors is zero, the term in the quark-quark potential induced by the Fourier transform of the constant potential and the divergence induced by the Fourier transform of the Coulomb and linear confinement potentials cancel out, respectively. The spin factor is
{\small{
\begin{eqnarray*}
&&\mathcal{I}_\text{spin}=\langle\left[\chi_{s_c}^{C}(13)\otimes\chi_{s_d}^{D}(245)\right]_{S^{'}}
|\hat V_s|\left[\chi_{s_a}^{A}(12)\otimes\chi^{B}_{s_b}(345)\right]_{S}\rangle,\nonumber\\
\end{eqnarray*}}}
where $S^{(')}$ is the total spin of the initial (final) state, $s$ denotes the spin of the four hadrons.~$\hat V_s$ is defined as the spin-spin interaction operator. In the quark model, the $\hat V_s$ is unitary for the Coulomb and linear confinement interactions, and $\mathbf{S}_{i}\cdot\mathbf{S}_{j}$ for the hyperfine potential.~The derivation of the spin factor $\mathcal{I}_{\text{spin}}$ is illustrated in Appendix \ref{Appendixsf}. Numerical results for the color-spin-flavor factors are collected in Table \ref{Ifactor}.

\renewcommand\tabcolsep{0.15cm}
\renewcommand{\arraystretch}{1.8}
\begin{table}[!htbp]
\caption{Color factor $\mathcal I_{\text{color}}$. Here, the sum of the color factors are $0$.}\label{tab:Icolor}
\begin{tabular}{cccc|cccc}
\toprule[1pt]
\multicolumn{4}{c|}{$\text{Prior}$} & \multicolumn{4}{c}{Post}\tabularnewline
\hline
$23$ & $24(5)$ & $13$ & $14(5)$ & $32(5)$ & $34(5)$ & $12$ & $14(15)$\tabularnewline
\hline
$d_{1}$ & $d_{2}$ & $\bar{d_{1}}$ & $\bar{d_{2}}$ & $d'_{1}$ & $d'_{2}$ & $\bar{d'_{1}}$ & $\bar{d'_{2}}$\tabularnewline
\hline
$\frac{4}{9}$ & $-\frac{2}{9}$ & $-\frac{4}{9}$ & $\frac{2}{9}$ & $\frac{2}{9}$ & $-\frac{2}{9}$ & $-\frac{4}{9}$ & $\frac{4}{9}$\tabularnewline
\bottomrule[1pt]
\end{tabular}
\end{table}

\renewcommand\tabcolsep{0.15cm}
\renewcommand{\arraystretch}{1.8}
\begin{table*}[!htbp]
\caption{Numerical results of the color-spin-flavor factors for different diagrams. The matrix elements in the last column are the weights for the $(d_1,d_2,\bar d_1,\bar d_2)$ and  $(d'_1,d'_2,\bar d'_1,\bar d'_2)$ diagrams.\label{Ifactor}}
\begin{tabular}{c|c|cccccc|c}
\toprule[1pt]
 & ($I$, $J$) & \multicolumn{6}{c|}{$\mathbf{S}_{i}\cdot\mathbf{S}_{j}$} & $\mathbf 1$\tabularnewline
\midrule[1pt]
 & ($\frac{1}{2},$ $\frac{1}{2}$) & $d_{1}$ & $d_{2}$ & $\bar{d}_{1}$ & $\bar{d_{2}}$ & $\bar{d}'_{1}$ & $d'_{2}$ &  Coulomb and linear \tabularnewline
\midrule[1pt]
\multirow{6}{*}{$J/\psi p$ } & $\Sigma_c^{++}D^{-}$ & $\frac{1}{18\sqrt{3}}$ & $-\frac{1}{9\sqrt{3}}$ & $\frac{1}{18\sqrt{3}}$ & $-\frac{1}{9\sqrt{3}}$ & $-\frac{1}{6\sqrt{3}}$ & $-\frac{1}{9\sqrt{3}}$ & $-\frac{2}{9\sqrt{3}}\{1,-1,-1,1\}$\tabularnewline
 & $\Sigma_c^{++}D^{*-}$ & $\frac{7}{54}$ & $\frac{5}{27}$ & $-\frac{5}{54}$ & $-\frac{1}{9}$ & $-\frac{5}{54}$ & $\frac{5}{27}$ & $\frac{10}{27}\{1,-1,-1,1\}$\tabularnewline
 & $\Sigma_c^{*++}D^{*-}$ & $-\frac{\sqrt{2}}{27}$ & $\frac{2\sqrt{2}}{27}$ & $-\frac{\sqrt{2}}{27}$ & $-\frac{\sqrt{2}}{9}$ & $-\frac{\sqrt{2}}{27}$ & $-\frac{\sqrt{2}}{27}$ & $\frac{4\sqrt{2}}{27}\{1,-1,-1,1\}$\tabularnewline
 & $\Sigma_c^{+}\bar{D}^0$ & $\frac{1}{18\sqrt{6}}$ & $-\frac{1}{9\sqrt{6}}$ & $\frac{1}{18\sqrt{6}}$ & $-\frac{1}{9\sqrt{6}}$ & $\frac{1}{6\sqrt{6}}$ & $-\frac{1}{9\sqrt{6}}$ &
 $-\frac{\sqrt{\frac{2}{3}}}{9} \{1,-1,-1,1\}$ \tabularnewline
 & $\Sigma_c^{+}\bar{D}^{*0}$ & $\frac{7}{54\sqrt{2}}$ & $\frac{5}{27\sqrt{2}}$ & $-\frac{5}{54\sqrt{2}}$ & $-\frac{1}{9\sqrt{2}}$ & $-\frac{5}{54\sqrt{2}}$ & $\frac{5}{27\sqrt{2}}$ & $\frac{5\sqrt{2}}{27}\{1,-1,-1,1\}$\tabularnewline
 & $\Sigma_c^{*+}\bar{D}^{*0}$ & $-\frac{1}{27}$ & $\frac{2}{27}$ & $-\frac{1}{27}$ & $-\frac{1}{9}$ & $-\frac{1}{27}$ & $-\frac{1}{27}$ & $\frac{4}{27}\{1,-1,-1,1\}$\tabularnewline
\midrule[1pt]
 & ($\frac{1}{2},$ $\frac{3}{2}$) & \multicolumn{6}{c|}{$\mathbf{S}_{i}\cdot\mathbf{S}_{j}$} & Coulomb and linear\tabularnewline
\midrule[1pt]
\multirow{6}{*}{$J/\psi p$ } & $\Sigma_c^{*++}D^{-}$ & $\frac{1}{9\sqrt{3}}$ & $-\frac{2}{9\sqrt{3}}$ & $\frac{1}{9\sqrt{3}}$ & $-\frac{2}{9\sqrt{3}}$ & $-\frac{1}{3\sqrt{3}}$ & $\frac{1}{9\sqrt{3}}$ & $-\frac{4}{9\sqrt{3}}\{1,-1,-1,1\}$\tabularnewline
 & $\Sigma_c^{++}D^{*-}$ & $\frac{5}{27}$ & $\frac{2}{27}$ & $-\frac{1}{27}$ & $0$ & $-\frac{1}{27}$ & $\frac{2}{27}$ & $\frac{4}{27}\{1,-1,-1,1\}$\tabularnewline
 & $\Sigma_c^{*++}D^{*-}$ & $-\frac{\sqrt{5}}{27}$ & $\frac{2\sqrt{5}}{27}$ & $-\frac{\sqrt{5}}{27}$ & $0$ & $-\frac{\sqrt{5}}{27}$ & $-\frac{\sqrt{5}}{27}$ & $\frac{4\sqrt{5}}{27}\{1,-1,-1,1\}$\tabularnewline
 & $\Sigma_c^{*+}\bar{D}^{0}$ & $\frac{1}{9\sqrt{6}}$ & $-\frac{\sqrt{\frac{2}{3}}}{9}$ & $\frac{1}{9\sqrt{6}}$ & $-\frac{\sqrt{\frac{2}{3}}}{9}$ & $-\frac{1}{3\sqrt{6}}$ & $\frac{1}{9\sqrt{6}}$ & $-\frac{1}{9}\left(2\sqrt{\frac{2}{3}}\right)\{1,-1,-1,1\}$\tabularnewline
 & $\Sigma_c^{+}\bar{D}^{*0}$ & $\frac{5}{27\sqrt{2}}$ & $\frac{\sqrt{2}}{27}$ & $-\frac{1}{27\sqrt{2}}$ & $0$ & $-\frac{1}{27\sqrt{2}}$ & $\frac{\sqrt{2}}{27}$ & $\frac{2\sqrt{2}}{27}\{1,-1,-1,1\}$\tabularnewline
 & $\Sigma_c^{*+}\bar{D}^{*}$ & $-\frac{\sqrt{\frac{5}{2}}}{27}$ & $\frac{\sqrt{10}}{27}$ & $-\frac{\sqrt{\frac{5}{2}}}{27}$ & $0$ & $-\frac{\sqrt{\frac{5}{2}}}{27}$ & $-\frac{\sqrt{\frac{5}{2}}}{27}$ & $\frac{2\sqrt{10}}{27}\{1,-1,-1,1\}$\tabularnewline
\midrule[1pt]
 & ($\frac{1}{2},$ $\frac{1}{2}$) & \multicolumn{6}{c|}{$\mathbf{S}_{i}\cdot\mathbf{S}_{j}$} & {Coulomb} and linear\tabularnewline
\midrule[1pt]
\multirow{6}{*}{$\eta_{c}p$ } & $\Sigma_c^{++}D^{-}$ & $\frac{1}{6}$ & $\frac{1}{9}$ & $\frac{1}{6}$ & $\frac{1}{9}$ & $\frac{1}{6}$ & $\frac{1}{9}$ & $\frac{2}{9}\{1,-1,-1,1\}$\tabularnewline
 & $\Sigma_c^{++}D^{*-}$ & $\frac{1}{18\sqrt{3}}$ & $-\frac{1}{9\sqrt{3}}$ & $-\frac{1}{6\sqrt{3}}$ & $-\frac{1}{9\sqrt{3}}$ & $\frac{1}{18\sqrt{3}}$ & $-\frac{1}{9\sqrt{3}}$ & $-\frac{2}{9\sqrt{3}}\{1,-1,-1,1\}$\tabularnewline
 & $\Sigma_c^{*++}D^{*-}$ & $-\frac{\sqrt{\frac{2}{3}}}{9}$ & $\frac{2\sqrt{\frac{2}{3}}}{9}$ & $\frac{\sqrt{\frac{2}{3}}}{3}$ & $-\frac{\sqrt{\frac{2}{3}}}{9}$ & $-\frac{\sqrt{\frac{2}{3}}}{9}$ & $-\frac{\sqrt{\frac{2}{3}}}{9}$ & $\frac{4\sqrt{\frac{2}{3}}}{9}\{1,-1,-1,1\}$\tabularnewline
 & $\Sigma_c^{+}\bar{D}^{0}$ & $\frac{1}{6\sqrt{2}}$ & $\frac{1}{9\sqrt{2}}$ & $\frac{1}{6\sqrt{2}}$ & $\frac{1}{9\sqrt{2}}$ & $\frac{1}{6\sqrt{2}}$ & $\frac{1}{9\sqrt{2}}$ & $\frac{\sqrt{2}}{9}\{1,-1,-1,1\}$\tabularnewline
 & $\Sigma_c^{+}\bar{D}^{*0}$ & $\frac{1}{18\sqrt{6}}$ & $-\frac{1}{9\sqrt{6}}$ & $-\frac{1}{6\sqrt{6}}$ & $-\frac{1}{9\sqrt{6}}$ & $\frac{1}{18\sqrt{6}}$ & $-\frac{1}{9\sqrt{6}}$ & $-\frac{\sqrt{\frac{2}{3}}}{9}\{1,-1,-1,1\}$\tabularnewline
 & $\Sigma_c^{*+}\bar{D}^{*}$ & $-\frac{1}{9\sqrt{3}}$ & $\frac{2}{9\sqrt{3}}$ & $\frac{1}{3\sqrt{3}}$ & $-\frac{1}{9\sqrt{3}}$ & $-\frac{1}{9\sqrt{3}}$ & $-\frac{1}{9\sqrt{3}}$ & $\frac{4}{9\sqrt{3}}\{1,-1,-1,1\}$\tabularnewline
\bottomrule[1pt]
\end{tabular}
\end{table*}

Additionally, the explicit forms of spatial factors $\mathcal{I}_{\text{space}}$ for different diagrams can be written as
\begin{widetext}
\begin{eqnarray}
\mathcal{I}^{d_{1}}_{\text{space}} & =&\int\int\int d{\mathbf q}d{\mathbf p_{3}}d\mathbf{p_{4}}\phi_{A}(\mathbf q-\mathbf{p_c}+f_{A}\mathbf{p}+\mathbf{p_3})
\phi_{B}(\mathbf{p_3},\mathbf{p_4},-\mathbf{p}-\mathbf{p_3}-\mathbf{p_4})V_{23}(q^2) \nonumber\\
&\times &\phi_{C}^{*}(\mathbf q+\mathbf{p_3}-f_C{\mathbf{p_c}})\phi_{D}^{*}(\mathbf p+ \mathbf{p_3} - \mathbf{p_c}, \mathbf{p_4}, -\mathbf p - \mathbf{p_3} -\mathbf{p_4} ),\\
\mathcal{I}^{d_{2}}_{\text{space}}  & =&\int\int\int d{\mathbf q}d{\mathbf p_{3}}d\mathbf{p_{4}}\phi_{A}(-\mathbf{p_c}+f_{A}\mathbf{p}+\mathbf{p_3})
\phi_{B}(\mathbf{p_3},\mathbf{p_4},-\mathbf{p}-\mathbf{p_3}-\mathbf{p_4})V_{24}(q^2)\nonumber \\
 &\times &\phi_{C}^{*}(\mathbf{p_3}-f_C{\mathbf{p_c}})\phi_{D}^{*}(\mathbf{p}+\mathbf{p_3}+\mathbf q-\mathbf{p_c}, \mathbf{p_4}-\mathbf q,-\mathbf{p}-\mathbf{p_3}-\mathbf{p_4}),\\
\mathcal{I}^{\bar{d}_{1}} _{\text{space}}& =&\int\int\int d{\mathbf q}d{\mathbf p_{3}}d\mathbf{p_{4}}\phi_{A}(\mathbf{p_4}-\mathbf{p_c}+f_{A}\mathbf{p})\phi_{B}
(\mathbf{p_3},\mathbf{p_4},-\mathbf{p}-\mathbf{p_3}-\mathbf{p_4})V_{13}(q^2)\nonumber\\
 &\times &\phi_{C}^{*}(\mathbf{p_3}-\mathbf q-f_C\mathbf{p_c})\phi_{D}^{*}
 (\mathbf{p}+\mathbf{p_3}-\mathbf{p_c},\mathbf{p_4},-\mathbf{p}-\mathbf{p_3}-\mathbf{p_4}),\\
\mathcal{I}^{\bar{d}_{2}}_{\text{space}}& =&\int\int\int d{\mathbf q}d{\mathbf p_{3}}d\mathbf{p_{4}}\phi_{A}(\mathbf{p_3}+\mathbf q-\mathbf{p_c}+f_{A}\mathbf{p})\phi_{B}
(\mathbf{p_3},\mathbf{p_4},-\mathbf{p}-\mathbf{p_3}-\mathbf{p_4})V_{14}(q^2)\nonumber\\
 & \times & \phi_{C}^{*}(\mathbf{p_3}-f_C{\mathbf{p_c}})\phi_{D}^{*}
 (\mathbf{p}+\mathbf{p_3}-\mathbf{p_c}+\mathbf q,\mathbf{p_4}-\mathbf q,-\mathbf{p}-\mathbf{p_3}-\mathbf{p_4}),\\
\mathcal{I}^{d'_{2}}_{\text{space}} & =&\int\int\int d{\mathbf q}d{\mathbf p_{3}}d\mathbf{p_{4}}\phi_{A}(\mathbf{p_3}+\mathbf q-\mathbf{p_c}+f_{A}\mathbf{p})\phi_{B}
(\mathbf{p_3},\mathbf{p_4},-\mathbf{p}-\mathbf{p_3}-\mathbf{p_4})V_{34}(q^2)\nonumber\\
 & \times & \phi_{C}^{*}(\mathbf{p_3}+\mathbf q-f_C{\mathbf{p_c}})\phi_{D}^{*}(\mathbf{p}+\mathbf{p_3}-\mathbf{p_c}+\mathbf q,\mathbf{p_4}-\mathbf q,-\mathbf{p}-\mathbf{p_3}-\mathbf{p_4}),\\
\mathcal{I}^{\bar d'_{1}}_{\text{space}} & =&\int\int\int d{\mathbf q}d{\mathbf p_{3}}d\mathbf{p_{4}}\phi_{A}(\mathbf{p_3}+\mathbf q-\mathbf{p_c}+f_{A}\mathbf{p})\phi_{B}
(\mathbf{p_3},\mathbf{p_4},-\mathbf{p}-\mathbf{p_3}-\mathbf{p_4})V_{12}(q^2)\nonumber\\
 & \times & \phi_{C}^{*}(\mathbf{p_3}-f_C{\mathbf{p_c}})\phi_{D}^{*}
 (\mathbf{p}+\mathbf{p_3}-\mathbf{p_c},\mathbf{p_4},-\mathbf{p}-\mathbf{p_3}-\mathbf{p_4}),
\end{eqnarray}
\end{widetext}
where the $\mathbf{p_{3(4)}}$ is defined as the momentum of the third (fourth) quark. $f_A$ and $f_C$ are expressed as
\begin{eqnarray}
f_A=\frac{{m_1}}{{m_1}+{m_2}}, \quad\quad\quad {f_C}=\frac{m_3}{m_1+m_3},
\end{eqnarray}
with $m_i$ being the mass of the $i$th quark. $\phi(\mathbf p)$ is the spatial wave function of the hadron presented in Appendix \ref{Appendixwf}. The integral of the linear confinement potential  is divergent at $q=0$. The divergency cancels out exactly with each other due to the color factors.
According to the isospin symmetry, one obtains the following relation:
\begin{eqnarray*}
&&t(D^{(*)-}\Sigma^{(*)++}_c\rightarrow J/\psi (\eta_c)p)=-\sqrt{2}t(\bar D^{(*)0}\Sigma^{(*)+}_c\rightarrow J/\psi(\eta_c) p).
\end{eqnarray*}
With the $T$-matrix, we obtain the effective potential $V_{\text{eff}}$ between the $D^{(*)}\Sigma^{(*)}_c$ and the $J/\psi(\eta_c) p$ channels.

\section{$P_c$ states as pure $\Sigma^{(*)}_c\bar D^{(*)}$ molecules}\label{sec3}

In scenario I, we study the decay patterns of the $P_c$ states as pure $\Sigma^{(*)}_c\bar D^{(*)}$ molecules. According to their mass spectra, the lowest $P_c(4312)$ is probability the $\Sigma_c \bar D$ molecule with $J^P=\frac{1}{2}^-$. The higher $P_c(4440)$ and $P_c(4457)$ are very likely to be the $\Sigma_c\bar D^*$ molecular states. Their $J^P$ quantum numbers are proposed to be $\frac{1}{2}^-$ and $\frac{3}{2}^-$ in Refs. \cite{ Chen:2019asm,Meng:2019ilv,Xiao:2019aya,He:2019ify,Liu:2019tjn}, whereas $\frac{3}{2}^-$ and $\frac{1}{2}^-$ in Refs. \cite{Yamaguchi:2019seo,Valderrama:2019chc,Liu:2019zvb}. Both of these two spin assignments will be discussed in this section.
\subsection{Heavy quark symmetry}
Before the numerical analysis, we would like to discuss the hidden-charm strong decay behaviors of $\Sigma^{(*)}_c\bar{D}^{(*)}$ molecular states with the heavy quark spin symmetry.~The corresponding interaction amplitudes of $\Sigma_c^{(*)}\bar D^{(*)}\rightarrow J/\psi (\eta_c) p$ processes are collected in Table \ref{HQS}. For the $\Sigma_c\bar{D}^{(*)}$ states with $J^P=\frac{1}{2}^-$, one obtains
\begin{eqnarray}
R_1 &=& \frac{\Gamma{\left(\Sigma_c\bar{D}[\frac{1}{2}^-]\to\eta_cp\right)}}
    {\Gamma{\left(\Sigma_c\bar{D}[\frac{1}{2}^-]\to J/\psi p\right)}}=3,\label{R1}\\
R_2 &=& \frac{\Gamma{\left(\Sigma_c\bar{D}^*[\frac{1}{2}^-]\to J/\psi p\right)}}
    {\Gamma{\left(\Sigma_c\bar{D}^*[\frac{1}{2}^-]\to \eta_cp\right)}}={25\over 3}.\label{R2}
\end{eqnarray}

\renewcommand\tabcolsep{0.15cm}
\renewcommand{\arraystretch}{1.8}
\begin{table}[!htbp]
\caption{The $T$-matrix of the process $\Sigma_c^{(*)}\bar D^{(*)}\rightarrow J/\psi p$ or $\eta_c p$ in the heavy quark limit. All the spatial information is included in the matrix element $H_{h,l}$. The subscripts $h$ and $l$ denote  the heavy and light degrees of freedom  in the initial and final states. }\label{HQS}
\begin{tabular}{ccc|cc}
\toprule[1pt]
$\frac{1}{2}^{-}$ & $\eta_{c}p$ & $J/\psi p$ & $\frac{3}{2}^{-}$ & $J/\psi p$\tabularnewline
\midrule[1pt]

$|\Sigma_{c}\bar{D}\rangle$ & $\frac{1}{2}H_{0,\frac{1}{2}}$ & $-\frac{1}{2\sqrt{3}}H_{1,\frac{1}{2}}$& $|\Sigma_{c}^{*}\bar D\rangle$ & $-\frac{1}{\sqrt{3}}H_{1,\frac{1}{2}}$\tabularnewline

$|\Sigma_{c}\bar{D}^{*}\rangle$ & $-\frac{1}{2\sqrt{3}}H_{0,\frac{1}{2}}$ & $\frac{5}{6}H_{1,\frac{1}{2}}$ & $|\Sigma_{c}\bar{D}^{*}\rangle$ & $\frac{1}{3}H_{1,\frac{1}{2}}$\tabularnewline

$|\Sigma_{c}^{*}\bar{D}^{*}\rangle$ & $\sqrt{\frac{2}{3}}H_{0,\frac{1}{2}}$ & $\frac{\sqrt{2}}{3}H_{1,\frac{1}{2}}$ & $|\Sigma_{c}^{*}\bar{D}^{*}\rangle$ & $\frac{\sqrt{5}}{3}H_{1,\frac{1}{2}}$\tabularnewline
\bottomrule[1pt]
\end{tabular}
\end{table}

For the $\Sigma_c\bar D^*$ state with $J^P=\frac{3}{2}^-$, its decay into the $\eta_cp$ and $J/\psi p$ processes occur via $D$-wave and $S$-wave interactions, respectively. {The decay width of $\Sigma_c\bar{D}^*[\frac{3}{2}^-]\to\eta_cp$ is strongly suppressed by a $(p_c/M)^4$ factor, which is of $\mathcal O(10^{-3})$}. Thus, we do not consider this decay process. Anyway, one can still find that
\begin{eqnarray}
R_3 &=& \frac{\Gamma{\left(\Sigma_c\bar{D}^*[\frac{1}{2}^-]\to J/\psi p\right)}}
    {\Gamma{\left(\Sigma_c\bar{D}^*[\frac{3}{2}^-]\to J/\psi p\right)}}={25\over 4}.\label{R3}
\end{eqnarray}
The above conclusions can also be applied to their partner $P_b$ states as the heavy quark flavor symmetry. According to Eqs. (\ref{R1}) and (\ref{R2}), we want to emphasize that the $\eta_cp$ final state is a very important decay channel to observe the $\Sigma_c\bar{D}$ bound states with $J^P=\frac{1}{2}^-$ because of its much larger decay ratio,{ whereas it is an unreasonable channel for the observation of the  $\Sigma_c\bar{D}^{*}$ states with $J^P=\frac{3}{2}^-$ due to the $D$-wave suppression. }

The near threshold behavior of $P_c$ states provides an intuitive explanation that they are good candidates for the hidden-charm meson-baryon molecules. The different $\Sigma^{(*)}_c\bar D^{(*)}$ configurations with the same $J^P$ may couple with one another and contribute to the same pentaquark state. In the following, we present the numerical results in two scenarios, which correspond to $P_c$ states as pure $\Sigma^{(*)}_c\bar D^{(*)}$ molecules and molecular admixtures, respectively.

\subsection{$P_c$ states as pure $\Sigma^{(*)}_c\bar D^{(*)}$ molecules}\label{s-I}

\begin{figure*}[!htbp]
\centering
\includegraphics[width=0.338\textwidth]{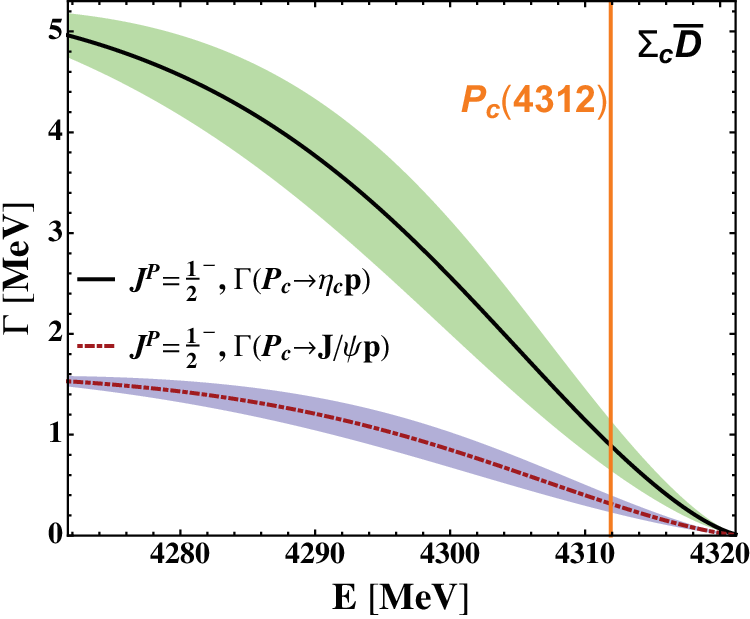}\,
\includegraphics[width=0.332\textwidth]{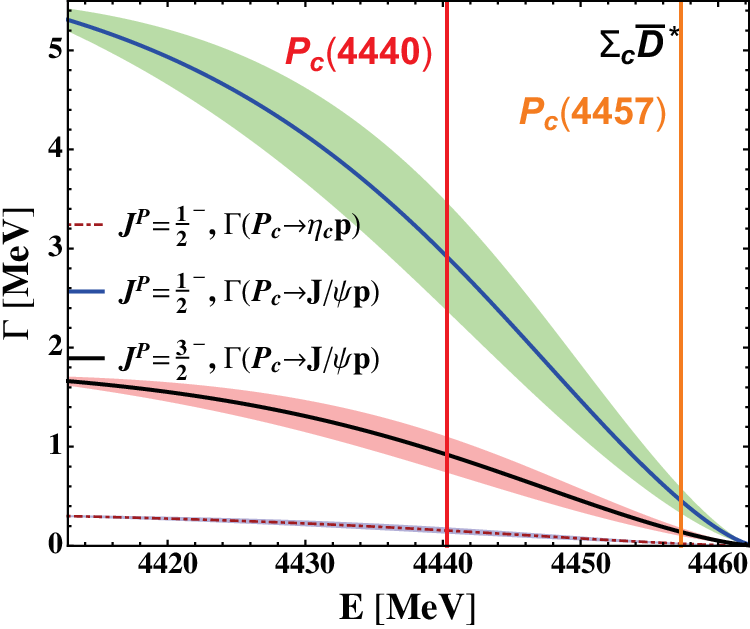}
\caption{ The mass dependence of the partial decay width for the $P_c$ states decaying into the $J/\psi p$ and $\eta_c p$ channels in scenario I. The binding energy is in the range of $ -50\sim -1$ MeV. }
\label{fig:massdenpendence}
\end{figure*}

\begin{figure*}[!htbp]
\centering
\includegraphics[width=0.35\textwidth]{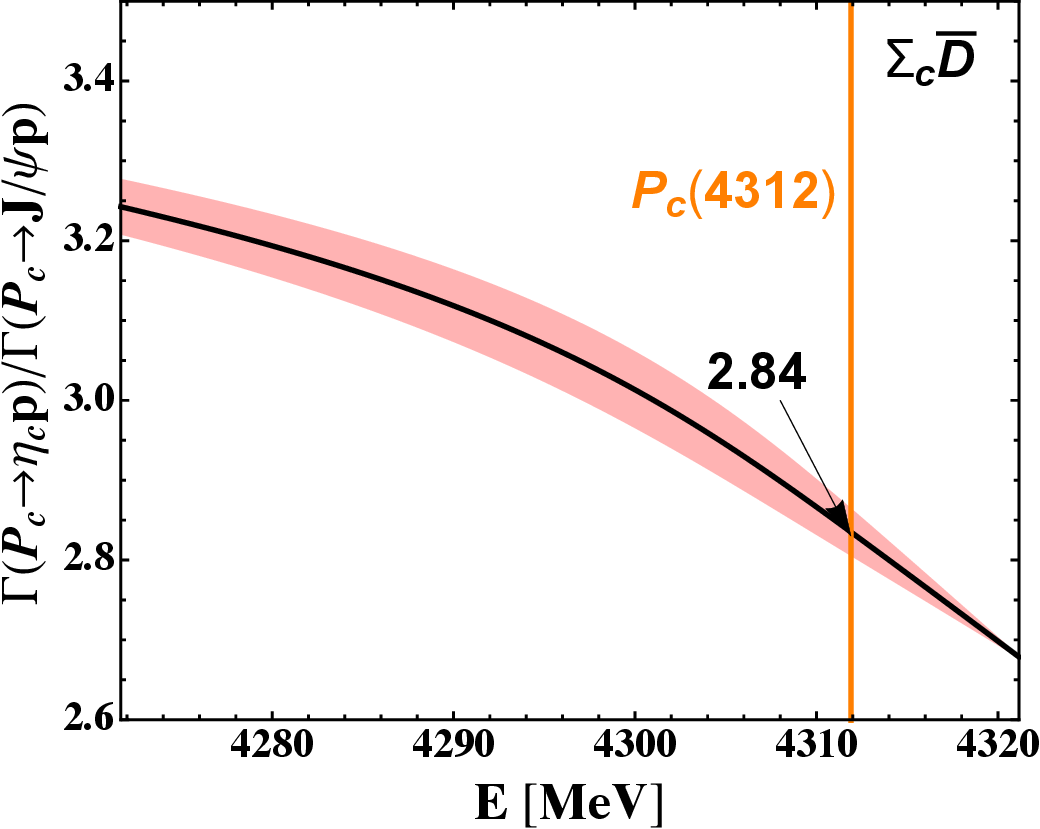}\,
\includegraphics[width=0.34\textwidth]{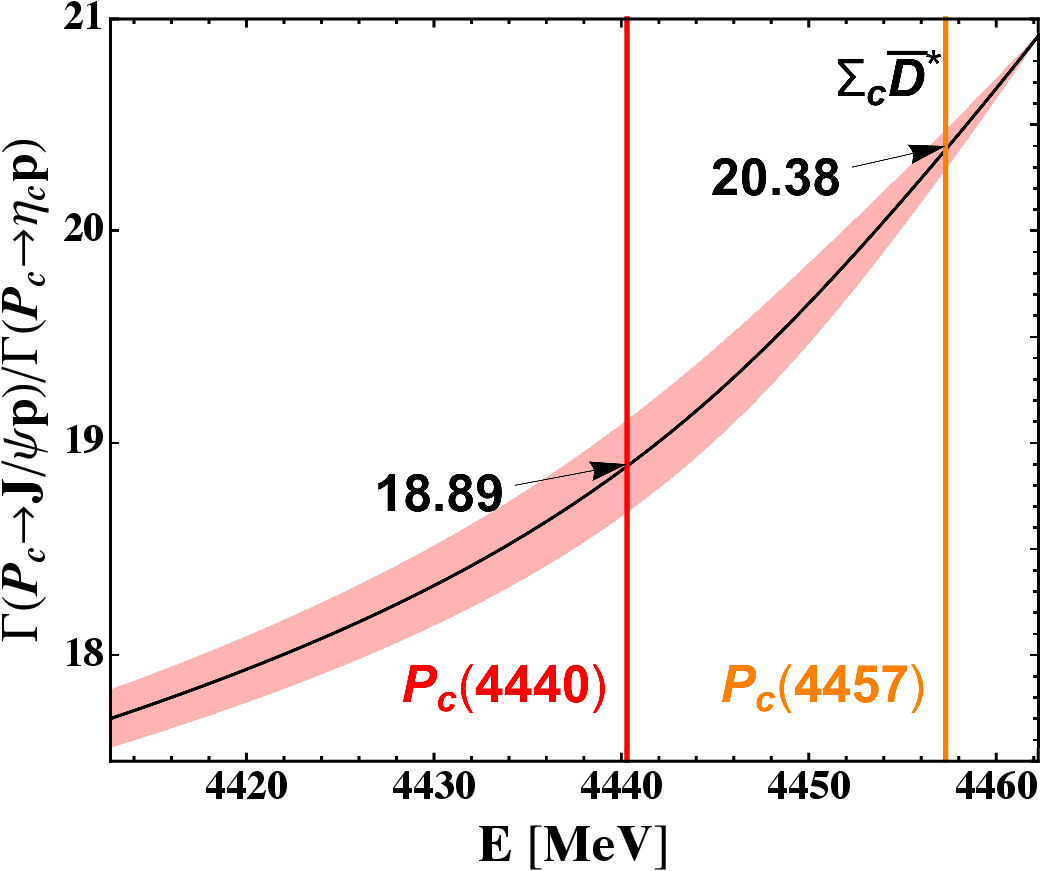}
\caption{ The branching fraction ratios for the $P_c$ states with $J^P=\frac{1}{2}^-$ decaying into the $J/\psi p$ and $\eta_c p$ channels. }
\label{fig:ratioC}
\end{figure*}

\begin{figure*}[!htbp]
\centering
\includegraphics[width=0.314\textwidth]{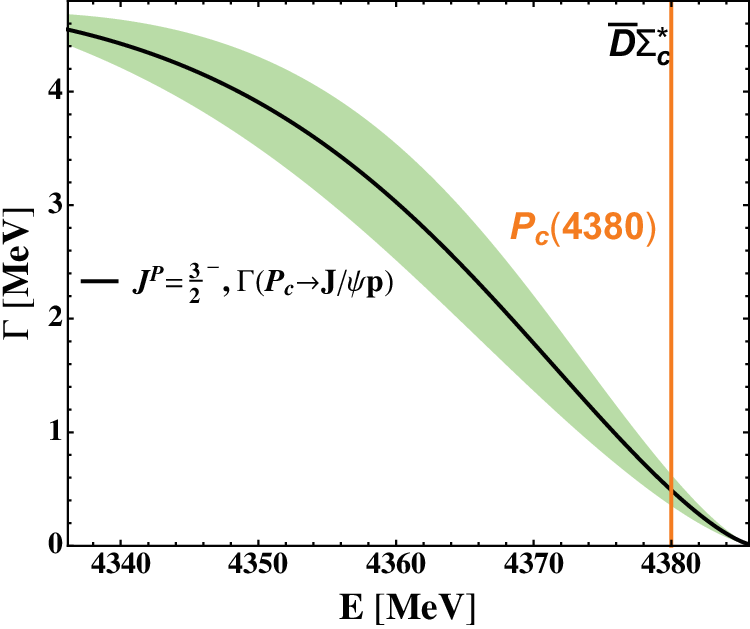}\quad
\includegraphics[width=0.32\textwidth]{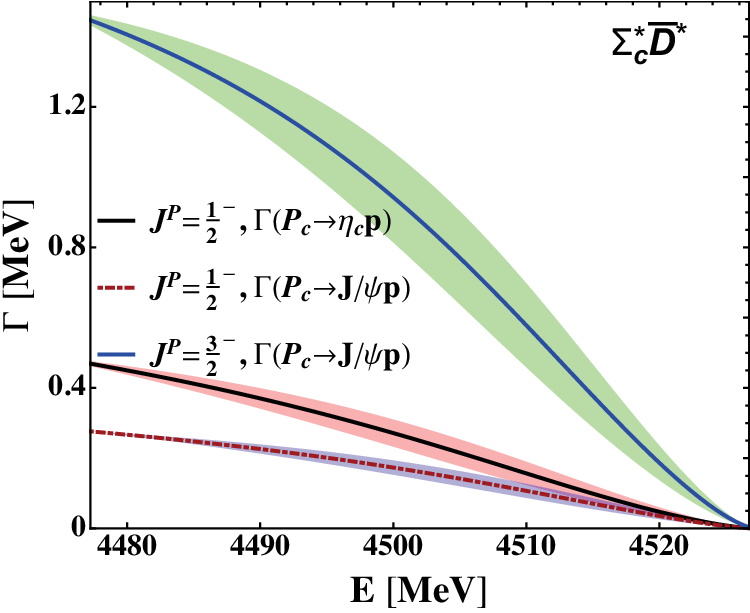}\quad
\includegraphics[width=0.32\textwidth]{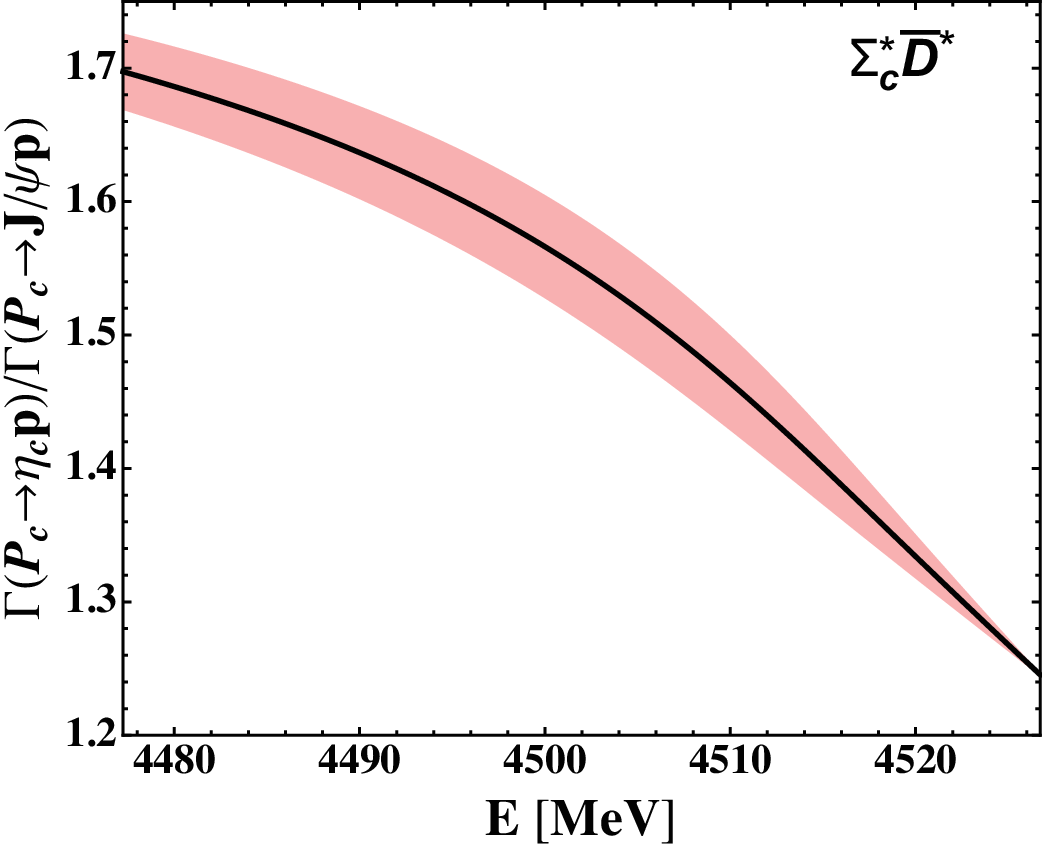}
\caption{ The mass dependence of the partial decay width for the $P_c$ states decaying into the $J/\psi p$ and $\eta_c p$ channels in scenario I. The binding energy is in the range of $-50\sim -1$ MeV. }
\label{fig:4380}
\end{figure*}

After the hidden-charmed strong decays discussed by the  heavy quark symmetry, we further perform a systematic analysis within the quark interchange model.   
Here, we adopt an S-wave Gaussian function with an undetermined oscillating parameter $\beta_{P_c}$ to estimate the relative wave function $R_{lS}(l=0)$ for the $P_c$ states in the meson-baryon picture.  The $\beta_{P_c}$ is related to the root mean square radius of the $P_c$ state. For an $S$-wave loosely bound molecule composed of two hadrons, the typical molecular size can be estimated as $r\sim1/{\sqrt{2\mu(M_A+M_B-M)}}$ with its reduced mass $\mu=\frac{M_AM_B}{M_A+M_B}$ \cite{Weinberg:1962hj,Weinberg:1963zza,Guo:2017jvc}. With this input, $\beta_{P_c} $ can be related to the mass of the $P_c$ state,
\begin{eqnarray}
\beta_{P_c}=\sqrt{3 \mu (M_A+M_B-M)}.
\end{eqnarray}
Moreover, we still allow $10\%$ uncertainty for this relation in the following numerical calculations. This leads to the uncertainties in Figs. \ref{fig:massdenpendence}-\ref{fig:4380}.

In Fig.~\ref{fig:massdenpendence}, we present the mass dependence of the partial decay width for the $\Sigma_c\bar D^{(*)}$ molecules decaying into the $J/\psi p$ and $\eta_cp$ channels. The binding energies for these $\Sigma_c\bar D^{(*)}$ molecules vary from $-50$ to $-1$ MeV. With a smaller binding energy, the $P_c$ state has a larger mass, which results in a larger relative momentum in the final state.~As illustrated in Eq.~(\ref{Tll}), the decay width depends on both the final momentum and the potential $V_\text{eff}(p_c, p, \mu)$. With the increasing initial and final relative momenta, the spatial factor $\mathcal I_{\text{space}}$ suffers an exponential suppression and the effective potential decreases. In the limit of large relative momentum of the final states, the $\mathcal I_{\text{space}}$ and the decay width vanish. As shown in Fig. \ref{fig:massdenpendence}, when the binding energy is taken as $-50$ MeV, the decay widths of $P_c\to J/\psi p (\eta_c p)$ become the largest with their smaller phase space.

In Fig.~\ref{fig:ratioC}, we present the decay ratio $\Gamma(P_c\rightarrow \eta_cp)/\Gamma(P_c\rightarrow J/\psi p)$. We find that the decay ratio decreases with the larger $P_c$ mass because the relative momentum in the $\eta_cp$ channel is larger than that in the $J/\psi p$ channel. The effective potential decreases faster for the $\eta_c p$ channel.

As shown in Fig.~\ref{fig:ratioC}, the loosely bound $\Sigma_c\bar D$ molecule prefers to decay into the  $\eta_cp$ channel rather than the $J/\psi p$ channel with its binding energy in the range of $-50\sim -1$ MeV. If the $P_c(4312)$ is the $\Sigma_c\bar D$ molecule, its partial widths decaying into the $\eta_c p$ and $J/\psi p$ channels are $0.89\pm0.25$ MeV and $0.32\pm 0.08$ MeV, respectively. Here, the errors come from the uncertainty of the relative molecular wave function. The decay ratio $R_1$ is $2.84\pm0.03$. The substantial reduction of a relative error in $R_1$ results from a strong correlation of theoretical uncertainties in individual partial widths considered. Thus, the $\eta_c p$ should be the other promising decay channel to observe the $P_c(4312)$ molecular state.

For the $\Sigma_c\bar D^*$ molecules with $J^P=\frac{1}{2}^-$ and $J^P=\frac{3}{2}^-$, as shown in Fig. \ref{fig:massdenpendence}, the $J/\psi p$ decay channel is remarkably more important than the $\eta_cp$ channel. Even with a larger phase space in the $\eta_cp$ channel, the $\Sigma_c\bar D^*$ state decays much more easily into $J/\psi p$. The $\Sigma_c\bar D^*$ couples more strongly with the $J/\psi p$ channel. For the $\Sigma_c\bar D^*$ molecule with $J^P=\frac{1}{2}^-$, the ratios $R_2$ in Eq.~(\ref{R2}) are $18.89\pm 0.22$ and $20.38\pm 0.09$ at $M=4440.3$ and $4457.3$ MeV, respectively. It is interesting to note that the partial decay width of the $J/\psi p$ mode for the $\Sigma_c\bar D^*$ molecule with $J^P=\frac{1}{2}^-$ is larger than that in the $J^P=\frac{3}{2}^-$ state. The interaction between $\Sigma_c\bar D^*$ and $J/\psi p$ is sensitive to the total angular momentum. The interaction between the $\Sigma_c\bar D^*$ with $J^{P}=\frac{1}{2}^-$ and the $J/\psi p$ channel is stronger than that of the $J^P=\frac{3}{2}^-$ one.

In our calculation, the decay width of $\Gamma(P_c(4440)[1/2^-]\to J/\psi p)$ is $2.92\pm0.55$ MeV and $0.92\pm 0.18$ MeV for the $\Gamma(P_c(4440)[3/2^-]\to J/\psi p)$. Their corresponding branch fractions\footnote{The central value of the $P_c$ decay width is used to estimate the branching fractions.} are
\begin{eqnarray}
\mathcal{B}[\frac{1}{2}^-] &=& \frac{\Gamma(P_c(4440)[1/2^-]\to J/\psi p)}{\Gamma_{\text{total}}(P_c(4440))}=14.2\%,\\
\mathcal{B}[\frac{3}{2}^-] &=& \frac{\Gamma(P_c(4440)[3/2^-]\to J/\psi p)}{\Gamma_{\text{total}}(P_c(4440))}=4.4\%,
\end{eqnarray}
respectively. When the mass of the $\Sigma_c\bar{D}^*$ bound states with $J^P=1/2^-(3/2^-)$ is fixed as 4457 MeV, the above decay widths become $\Gamma'[1/2^-]=0.45\pm0.13$ and $\Gamma'[3/2^-]=0.14\pm 0.04$ MeV, respectively. Then, the branch fractions for $P_c(4457)$ with different spin parity are $\mathcal{B}'[1/2^-]=7.0\%$ and $\mathcal{B}'[3/2^-]=2.2\%$. By assuming $J^P=\frac{3}{2}^-$ for both $P_c$ states, the $\mathcal{B}[\frac{3}{2}^-] $ and $\mathcal{B'}[\frac{3}{2}^-] $ are $4.4\%$ and $2.2\%$, of which the upper limits are determined to be $2.3\%$ and $3.8\%$ at $90\%$ confidence level in  GlueX \cite{Ali:2019lzf}.

As a by-product, we also extend our calculations to the $\Sigma_c^*\bar{D}$ state with $J^P=3/2^-$ and $\Sigma_c^*\bar{D}^*$ states with $J^P=1/2^-, 3/2^-$, which are possible meson-baryon molecular candidates, see, e.g., Refs.~\cite{Liu:2019tjn,Xiao:2019aya,Sakai:2019qph}. We illustrate the mass dependence of the partial decay widths and the decay ratios for the $\Sigma_c^*\bar D^*$ molecules with $J^P=1/2^-$ in Fig.~\ref{fig:4380}.

The numerical results show that
\begin{enumerate}
  \item If the $P_c(4380)$ is the $\Sigma^*_c\bar D$ molecule, the partial decay width for the $P_c(4380)\to J/\psi p$ is $0.49\pm0.14$ MeV.
  \item The $\Sigma_c^*\bar{D}^*$ molecular state with $J^P=1/2^-$ prefers to decay into the $\eta_c p$ channel rather than the $J/\psi p$ channel.
  \item For the $\Sigma_c^*\bar{D}^*$ molecular state in $J^{P}=\frac{3}{2}^-$, it couples much more strongly with the $J/\psi p$ channel than that in the $J^P=\frac{1}{2}^-$ state.
  \item If the $\Sigma_c^*\bar D^*$ molecule with $J^P=\frac{3}{2}^-$ exists, its $\eta_c p$ decay mode is severely suppressed by the $D$-wave decay mechanism [of $\mathcal O(10^{-3})$].~Thus, it is much easier to be detected in the $J/\psi p$ channel.
\end{enumerate}

Motivated by the heavy quark flavor symmetry, the $P_b$ states, which are the corresponding states of the $P_c$ states, are excepted to be existing in the bottom sector \cite{Karliner:2015voa,Wu:2010rv,Cao:2019gqo,Wang:2019ato}. In our work, the study for the hidden-charm pentaquark states can be extended to the hidden-bottom case, once the wave functions of the pentaquark states are replaced by the bottom ones. In scenario I, we treat the $P_b$ states as pure $\Sigma^{(*)}_b B^{(*)}$ states and use an S-wave Gaussian function to mimic the relative molecular wave function. We perform the numerical calculations and list the results in Appendix \ref{Appendixpb}.

\section{$P_c$ states as admixtures of $\Sigma^{(*)}_c\bar D^{(*)}$ configurations}\label{sec4}

\begin{figure*}[!htbp]
\centering
\includegraphics[width=0.35\textwidth]{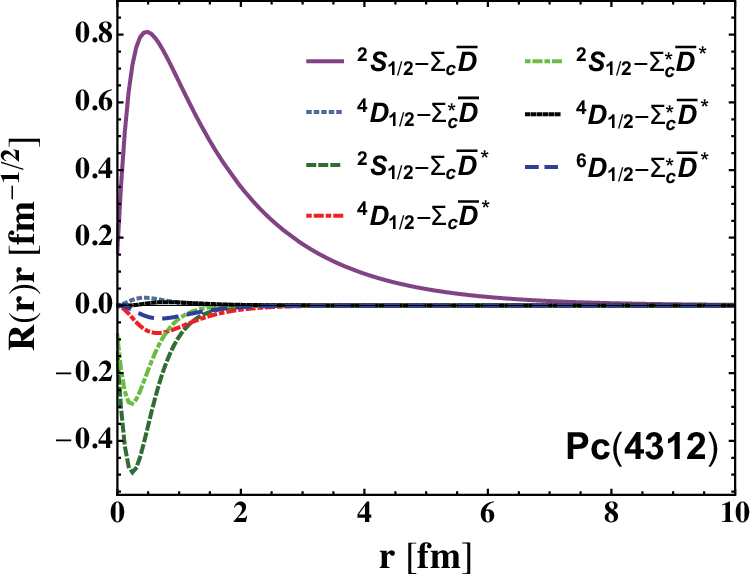}
\includegraphics[width=0.348\textwidth]{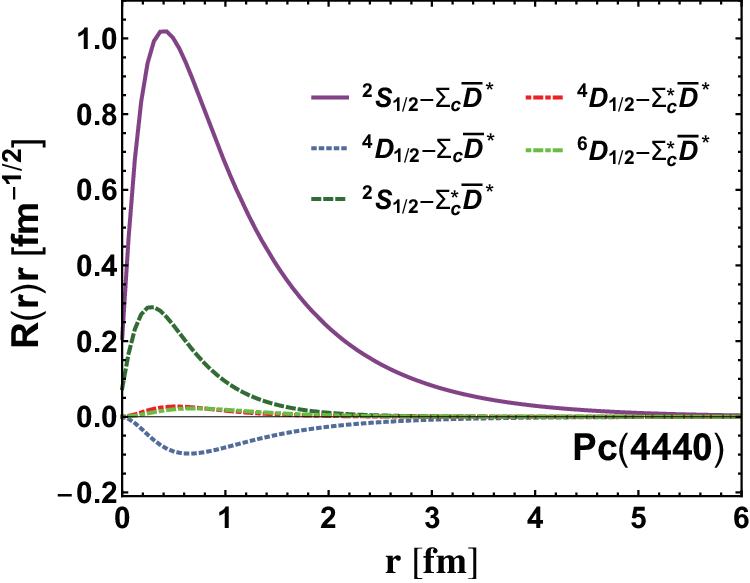}\\
\includegraphics[width=0.35\textwidth]{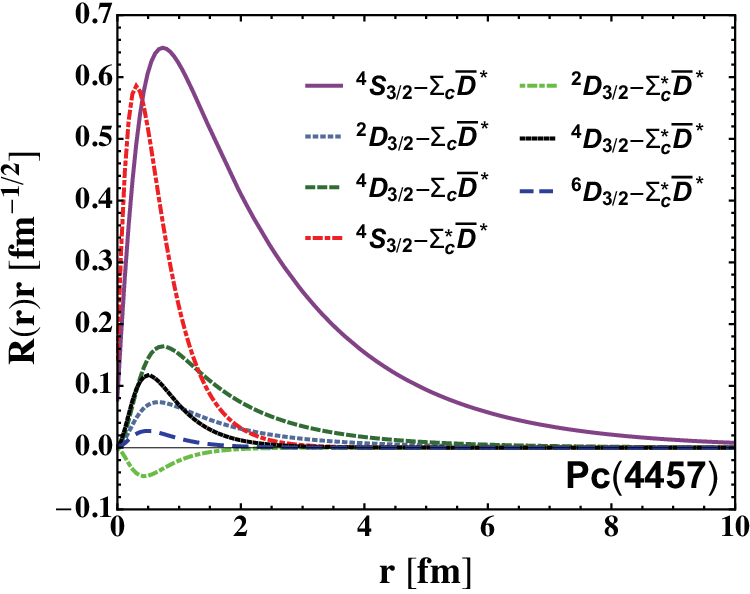}
\includegraphics[width=0.35\textwidth]{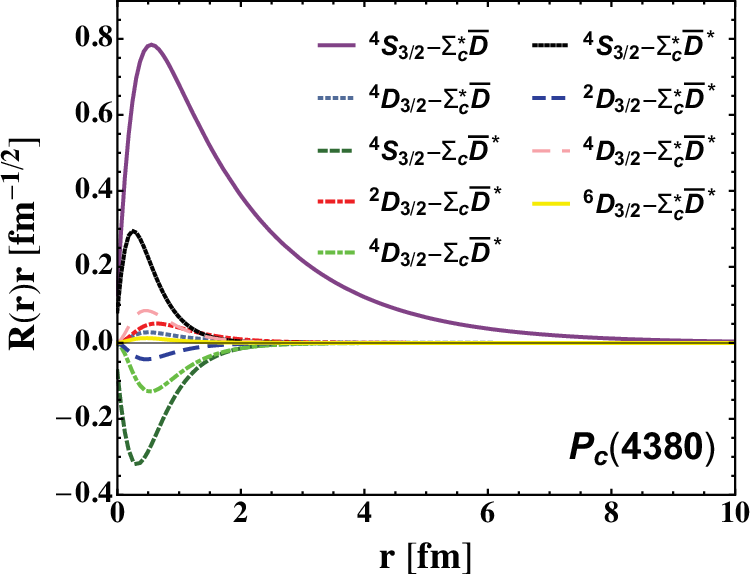}
\caption{ The relative wave functions between the two constituent hadrons in the molecular $P_c$ states in the OBE model. }
\label{fig:pc}
\end{figure*}

In scenario II,  we improve our results with the exact molecular wave functions obtained by solving the coupled channel Sch\"odinger equation in the scheme of the one-boson-exchange (OBE) model. The explicit details of the calculations are referred to Ref. \cite{Chen:2019asm}. The relative molecular wave functions between the constituent meson and baryon are presented in Fig.~\ref{fig:pc}. In Ref. \cite{Chen:2019asm}, the $P_c$ states observed by the LHCb collaboration can coexist as the admixtures of $\Sigma^{(*)}_c\bar D^{(*)}$ molecular states as illustrated in Table \ref{tab:pc}. The $P_c(4312)$ and $P_c(4440)$ are the $J^P=\frac{1}{2}^-$ molecular states mainly composed of $\Sigma_c\bar D$ and $\Sigma_c\bar D^*$ channels, while the $P_c(4457)$ and $P_c(4380)$ are the $J^P=\frac{3}{2}^-$ molecules mainly composed of $\Sigma_c\bar D^*$ and $\Sigma^*_c\bar D$ channels, respectively. In particular, the other flavor configurations also provide important contributions to reproduce the $P_c$ states. For instance, the $\Sigma^*_c\bar D^*$ channel couples with the $\Sigma_c\bar D^*$ and contributes to the $P_c(4457)$ state with its probability around $21\%$. Since the $D$-wave components contribute a tiny proportion to the $P_c$ states with their probabilities less than $5\%$, the decay from the $D$-wave components into the $J/\psi p$ ($\eta_cp$) channels will be suppressed. Here, we only consider the $S$-wave flavor configurations in calculating the partial decay widths.

\renewcommand\tabcolsep{0.11cm}
\renewcommand{\arraystretch}{1.8}
\begin{table*}
\caption{The masses  (in unit of MeV) of the hidden-charm $P_c$ states and probabilities for different flavor configurations in the states \cite{Chen:2019asm}. The symbol ``$\sim 0\% $" denotes that the proportion of the  corresponding configuration is tiny and negligible.}\label{tab:pc}
\begin{tabular}{c|ccccccccc}
\toprule[1pt]
 \multicolumn{10}{c}{$P_{c}(4312)$ $I(J^{P})=\frac{1}{2}(\frac{1}{2}^{-})$}\tabularnewline
\hline
Mass & $\Sigma_{c}\bar{D}|^{2}S_{\frac{1}{2}}\rangle$ & $\Sigma_{c}^{*}\bar{D}|^{4}D_{\frac{1}{2}}\rangle$ & $\Sigma_{c}\bar{D}^{*}|^{2}S_{\frac{1}{2}}\rangle$ & $\Sigma_{c}\bar{D}^{*}|^{4}D_{\frac{1}{2}}\rangle$ & $\Sigma_{c}^{*}\bar{D}^{*}|^{2}S_{\frac{1}{2}}\rangle$ & $\Sigma_{c}^{*}\bar{D}^{*}|^{4}D_{\frac{1}{2}}\rangle$ & $\Sigma_{c}^{*}\bar{D}^{*}|^{6}D_{\frac{1}{2}}\rangle$ & \tabularnewline
 4312.75 & {$84\%$} & $\sim0\%$ & $11\%$ & $\sim1\%$ & $4\%$ & $\sim0\%$ & $\sim0\%$ & \tabularnewline
\midrule[1pt]
\multicolumn{10}{c}{$P_{c}(4440)$ $I(J^{P})=\frac{1}{2}(\frac{1}{2}^{-})$}\tabularnewline
\hline
Mass &$\Sigma_{c}\bar{D}^{*}|{}^{2}S_{\frac{1}{2}}\rangle$  & $\Sigma_{c}\bar{D}^{*}|{}^{4}D_{\frac{1}{2}}\rangle$ & $\Sigma_{c}^{*}\bar{D}^{*}|{}^{2}S_{\frac{1}{2}}\rangle$ & $\Sigma_{c}^{*}\bar{D}^{*}|{}^{4}D_{\frac{1}{2}}\rangle$ & $\Sigma_{c}^{*}\bar{D}^{*}|{}^{6}D_{\frac{1}{2}}\rangle$ &  & \tabularnewline

 4442.88 & $94\%$ & $1\%$ & $5\%$ & $\sim0\%$ & $\sim0\%$ &  &  &  & \tabularnewline
\midrule[1pt]
\multicolumn{10}{c}{$P_{c}(4457)$ $I(J^{P})=\frac{1}{2}(\frac{3}{2}^{-})$}\tabularnewline
\hline
Mass &  $\Sigma_{c}\bar{D}^{*}|^{4}S_{\frac{3}{2}}\rangle$ & $\Sigma_{c}\bar{D}^{*}|{}^{4}D_{\frac{3}{2}}\rangle$ & $\Sigma_{c}\bar{D}^{*}|^{4}S_{\frac{3}{2}}\rangle$ & $\Sigma_{c}^{*}\bar{D}^{*}|^{4}S_{\frac{3}{2}}\rangle$ & $\Sigma_{c}^{*}\bar{D}^{*}|^{2}D_{\frac{3}{2}}\rangle$ & $\Sigma_{c}^{*}\bar{D}^{*}|^{4}D_{\frac{3}{2}}\rangle$ & $\Sigma_{c}^{*}\bar{D}^{*}|^{6}D_{\frac{3}{2}}\rangle$ & \tabularnewline
4457.77 & $74\%$ & $1\%$ & $3\%$ & $21\%$ & $\sim0\%$ & $1\%$ & $\sim0\%$ & \tabularnewline
\midrule[1pt]
\multicolumn{10}{c}{$P_{c}(4380)$ $I(J^{P})=\frac{1}{2}(\frac{3}{2}^{-})$}\tabularnewline
\hline
Mass & $\Sigma_{c}^{*}\bar{D}|{}^{4}S_{\frac{3}{2}}\rangle$ & $\Sigma_{c}^{*}\bar{D}|{}^{4}D_{\frac{3}{2}}\rangle$ & $\Sigma_{c}\bar{D}^{*}|{}^{4}S_{\frac{3}{2}}\rangle$ & $\Sigma_{c}\bar{D}^{*}|{}^{2}D_{\frac{3}{2}}\rangle$ & $\Sigma_{c}\bar{D}^{*}|{}^{4}D_{\frac{3}{2}}\rangle$ & $\Sigma_{c}^{*}\bar{D}^{*}|{}^{4}S_{\frac{3}{2}}\rangle$ & $\Sigma_{c}^{*}\bar{D}^{*}|{}^{2}D_{\frac{3}{2}}\rangle$ & $\Sigma_{c}^{*}\bar{D}^{*}|{}^{4}D_{\frac{3}{2}}\rangle$ & $\Sigma_{c}^{*}\bar{D}^{*}|{}^{6}D_{\frac{3}{2}}\rangle$\tabularnewline
4379.11 & $87\%$ & $\sim0\%$ & $6\%$ & $\sim0\%$ & $1\%$ & $4\%$ & $\sim0\%$ & $\sim0\%$ & $\sim0\%$\tabularnewline
\bottomrule[1pt]
\end{tabular}
\end{table*}

\renewcommand\tabcolsep{0.15cm}
\renewcommand{\arraystretch}{1.8}
\begin{table}
\caption{The matrix elements $T_{ll}$  for the decays $P_c \rightarrow J/\psi p$ and $P_c \rightarrow \eta_c p$ with the $P_c$ as the admixture of different $\Sigma^{(*)}_c\bar D^{(*)}$ molecular states. The ``PW'' denotes the partial decay width.\label{cpresult}}
\begin{tabular}{c|c|ccc|c}
\toprule[1pt]\toprule[1pt]
\multicolumn{2}{c|}{\text{Scenario II}} & \multicolumn{3}{c|}{$T_{00}$} & \tabularnewline
 \midrule[1pt]
\multirow{6}{*}{$\frac{1}{2}^{-}$} & $P_{c}(4312)$ & $\Sigma_{c}\bar D$ & $\Sigma_{c}\bar D^{*}$ & $\Sigma_{c}^{*}\bar D^{*}$ & PW {[}MeV{]} \tabularnewline
\cline{2-6}
 & $\eta_{c}p$  & $-0.86$ & $-0.14$ & $0.16$ & \multirow{1}{*}{$1.98$}\tabularnewline

 & $J/\psi p$  & $0.55$ & $0.62$ & $0.13$ & \multirow{1}{*}{$3.77$}\tabularnewline
\cline{2-6}
 & $P_{c}(4440)$ & $\Sigma_{c}\bar D^{*}$ & $\Sigma_{c}^{*}\bar D^{*}$ &  & PW {[}MeV{]}\tabularnewline
\cline{2-6}
 & $\eta_{c}p$  & $-0.23$ & $0.09$ &  & $0.06$\tabularnewline

 & $J/\psi p$  & $1.06$ & $0.08$ &  & $3.79$\tabularnewline
\hline
\multirow{4}{*}{$\frac{3}{2}^{-}$} & $P_{c}(4457)$ & $\Sigma_{c}\bar D^{*}$ & $\Sigma_{c}^{*}\bar D^{*}$ &  & PW {[}MeV{]}\tabularnewline
\cline{2-6}
 & $J/\psi p$  & $-0.36$ & $-0.18$ &  & $0.90$\tabularnewline
\cline{2-6}
 & $P_{c}(4380)$ & $\Sigma_{c}^{*}\bar{D}$ & $\Sigma_{c}\bar{D}^{*}$ & $\Sigma_{c}^{*}\bar{D}^{*}$ & PW {[}MeV{]} \tabularnewline
\cline{2-6}
 & $J/\psi p$  & $0.82$ & $0.21$ & $-0.08$ & \multirow{1}{*}{$2.34$}\tabularnewline
\bottomrule[1pt]\bottomrule[1pt]
\end{tabular}
\end{table}

In Table \ref{cpresult}, we collect the $T$-matrices for the possible flavor configurations decaying into the $J/\psi p(\eta_cp)$ channels. Their sum contributes to the partial decay widths. It is interesting that the coupled channel effect significantly changes the decay pattern of the $P_c(4312)$. As illustrated in Sec.~\ref{s-I}, taking the $P_c(4312)$ as the pure $\Sigma_c \bar D$ molecule, the partial decay width ratio between the $\eta_cp$ and $J/\psi p$ is nearly $3$.~Considering the coupled channel effect, the $S$-wave component $\Sigma_c\bar{D}^*$ is considerable and occupies around $11\%$. However, its contribution to the $J/\psi p$ decay mode is larger than that from the dominant $\Sigma_c \bar D$ configuration and enlarges the partial decay width. These results indicate that the interaction between the $ \Sigma_c \bar D^*$ and $J/\psi p$ is much stronger than that in the $ \Sigma_c \bar D$ system with $J^P=1/2^-$. This is also consistent with the prediction in the heavy quark limit, in which the $V_\text{eff}( \Sigma_c \bar D^*-J/\psi p)$ is the largest while the $V_\text{eff}( \Sigma_c \bar D-J/\psi p)$ is the smallest for the channels with $J^P=\frac{1}{2}^-$. At present, $R_1$ becomes 0.53; this is strongly contrasted with the single channel ratio $2.84$. When we exclude the contribution from the $\Sigma_c \bar D^*$ configuration, the decay ratio $R_1$ of the $P_c(4312)$ becomes around $3$, which is similar to the value in scenario I.

In Ref. \cite{Chen:2019asm}, the coupled channel effect is also considerably large for the $P_c(4457)$. It contains $21\%$ $\Sigma^*_c\bar D^*$ configuration, which strongly couples with the $J/\psi p$ channel as illustrated in Sec.~\ref{s-I}. The interference effect between the contributions from the $\Sigma^*_c\bar D^*$ and $\Sigma_c\bar D^*$ is constructive and enlarges the partial decay width. Here, the decay width of $\Gamma(P_c(4457)\to J/\psi p)$ is 0.90 MeV, which is several times larger than that of the pure $\Sigma_c\bar{D}^*$ molecular state with $J^P=3/2^-$.

For the $P_c(4440)$ molecular state mainly composed of  the $\Sigma_c\bar{D}^*$ channel with $J^P=1/2^-$, as listed in Table \ref{cpresult}, the signs of the scattering amplitudes for the $\Sigma_c\bar D^*$ and $\Sigma^*_c\bar D^*$ with the $\eta_cp$ channel are opposite. The contributions partly cancel with each other. Finally, this cancellation leads to a quiet small partial decay width of the decay $P_c(4440)\rightarrow \eta_c p$, which is suppressed {by several orders} compared with the $J/\psi p$ decay. In other words, it may be a little hard to observe the $P_c(4440)$ in the $\eta_c p$ channel.

Since a loosely bound molecular state $P_c(4380)$ mainly composed of an $S$-wave $\Sigma_c^*\bar{D}$ component can be reproduced using the same set of parameters in our previous work \cite{Chen:2019asm}, we also obtain the decay width of $P_c(4380)\to J/\psi p$, $\Gamma=2.34$ MeV. The partial decay width of $P_c(4380)\to \eta_cp$ is suppressed by $\mathcal O(10^{-3})$.

\section{Summary} \label{sec5}

Inspired by the observations from the LHCb collaboration \cite{Aaij:2019vzc}, we have calculated the partial decay widths for the $P_c$ states as the $\Sigma^{(*)}_c\bar D^{(*)}$ molecules decaying into the $J/\psi p$ and $\eta_c p$ channels in the quark interchange model. Their partial decay widths are related with the scattering amplitudes between the $\Sigma^{(*)}_c\bar D^{(*)}$ and $J/\psi p$ ($\eta_c p$) channels, which are derived by the quark interchange model.~In the quark level, the interactions  between the hadrons are equivalently represented in terms of the interactions between the quarks. All the parameters in the quark model are determined by the mass spectra of the mesons.

In our calculations, we discuss the hidden-charm partial decay behaviors of the $P_c$ states as the pure (scenario I) or the coupled $\Sigma^{(*)}_c\bar D^{(*)}$ molecules (scenario II). The corresponding results are summarized in Table \ref{results}.

As a pure $\Sigma_c\bar D$ molecule with $J^P=1/2^-$, the $P_c(4312)$ has a larger decay width for the $\eta_c p$ decay mode than the $J/\psi p$ mode. Thus, one can expect the observation of the $P_c(4312)$ in the $\eta_c p$ decay channel. For the $P_c(4440)$/$P_c(4457)$ states in scenario I, the decay widths of $P_c\to J/\psi p$ are larger than that of $P_c\to \eta_c p$  {because of the stronger interaction between the $\Sigma_c\bar{D}^*$ state and the $J/\psi p$ channel.}

\renewcommand\tabcolsep{0.3cm}
\renewcommand{\arraystretch}{1.8}
\begin{table}[!htbp]
\caption{Comparison of the partial decay widths for the decay modes  $P_c\rightarrow J/\eta_cp$ and $P_c\rightarrow J/\psi p$ (in unit of MeV) as the pure $\Sigma^{(*)}_c \bar D ^{(*)}$ molecules (S-I) and their admixtures (S-II), respectively. The ``. . ." represent the absence of the corresponding results.  \label{results}}
{
\begin{tabular}{c|c|cc}
\toprule[1pt]
\multirow{1}{*}{ $J^P$} & \multirow{1}{*}{Channel} &  S-I & S-II \\
\midrule[1pt]

\multirow{3}{*}{$\frac{1}{2}^{-}$} & ${P_{c}(4312)\rightarrow\eta_{c}p}$ & $0.89\pm0.25$ & $1.98$ \tabularnewline

 & $P_{c}(4312)\rightarrow J/\psi p$ & $0.32\pm0.08$ & $3.77$  \tabularnewline

 & $\frac{P_{c}(4312)\rightarrow\eta_{c}p}{P_{c}(4312)\rightarrow J/\psi p}$ & $2.84\pm0.03$ & $0.53$  \tabularnewline
\hline
\multirow{3}{*}{$\frac{1}{2}^{-}$}  & $P_{c}(4440)\rightarrow\eta_{c}p$ & $0.15\pm0.03$ & $0.06$ \tabularnewline

 & $P_{c}(4440)\rightarrow J/\psi p$ & $2.92\pm0.55$ & $3.79$  \tabularnewline

 & $\frac{P_{c}(4440)\rightarrow J/\psi p}{P_{c}(4440)\rightarrow\eta_{c}p}$ & $18.89\pm0.22$ & $63.17$  \tabularnewline
\hline
$\frac{3}{2}^{-}$ & $P_{c}(4440)\rightarrow J/\psi p$ & $0.92\pm0.18$ & . . .  \tabularnewline
\hline
\multirow{3}{*}{$\frac{1}{2}^{-}$} & $P_{c}(4457)\rightarrow\eta_{c}p$ & $0.02\pm 0.01$ & . . . \tabularnewline

 & $P_{c}(4457)\rightarrow J/\psi p$ & $0.45\pm0.13$ & . . . \tabularnewline

 & $\frac{P_{c}(4457)\rightarrow J/\psi p}{P_{c}(4457)\rightarrow\eta_{c}p}$ & $20.38\pm0.09$  & . . . \tabularnewline
\hline
$\frac{3}{2}^{-}$ & $P_{c}(4457)\rightarrow J/\psi p$ & $0.14\pm0.04$ & $0.90$  \tabularnewline
\hline
$\frac{3}{2}^{-}$ & $P_{c}(4380)\rightarrow J/\psi p$ & $0.49\pm0.14$ & $2.34$  \tabularnewline
\bottomrule[1pt]
\end{tabular}
}
\end{table}

With the coupled channel effect between different flavor configurations taken into account, one can reproduce the masses of $P_c(4312)$, $P_c (4440)$, $P_c(4457)$, and $P_c(4380)$ states simultaneously in Ref. \cite{Chen:2019asm}. They are the molecules mainly composed of $\Sigma_c\bar D$ with $J^P=\frac{1}{2}^-$, $\Sigma_c\bar D^*$ with $J^P=\frac{1}{2}^-$ and $J^P=\frac{3}{2}^-$, and $\Sigma^*_c\bar D $ with $J^P=\frac{3}{2}^-$, respectively. Besides, the four $P_c$ states also contain other non-negligible flavor configurations. The coupled channel effect significantly influences the $J/\psi p$ decay mode of the $P_c(4312)$ and the $\eta_cp$ decay mode of the $P_c(4440)$. Our results indicate that the $P_c(4312)$ is easier to decay into the $J/\psi p$ than the $\eta_cp$ channel. Since the partial decay widths into the two channels are not so different, we expect $P_c(4312)$ can be observed in the $\eta_c p$ channel in the near future experiment, like the LHCb collaboration. The possibility of the observation of the $P_c$ states in the $\eta_c p$ mode is relevant to the relative contributions ($\mathcal R$ values) of the three narrow $P_c$ states, which are defined as 
\begin{eqnarray}
\mathcal R_{\eta_c}=\mathcal B(\Lambda_b \rightarrow P^+_cK^-)\mathcal B(P^+_c\rightarrow \eta_c p)/\mathcal B(\Lambda_b\rightarrow \eta_c p K^-).
\end{eqnarray}
The $\mathcal R_{\eta_c}$ is related to the relative contribution in the $J/\psi p$ mode by 
\begin{eqnarray}
\mathcal R_{\eta_c}=\mathcal R_{J/\psi_c}\times {{\mathcal B(P^+_c\rightarrow \eta_c p)}\over{\mathcal B(P^+_c\rightarrow J/ \psi p)}}\times
{{\mathcal B(\Lambda_b\rightarrow J/\psi p K^-)}\over{\mathcal B(\Lambda_b\rightarrow \eta_c p K^-)}}.
\end{eqnarray}
 In our work, we have calculated the ratio ${{\mathcal B(P^+_c\rightarrow \eta_c p)}\over{\mathcal B(P^+_c\rightarrow J/ \psi p)}}$.
The other ratio reads
\begin{eqnarray}
{{\mathcal B(\Lambda_b\rightarrow J/\psi p K^-)}\over{\mathcal B(\Lambda_b\rightarrow \eta_c p K^-)}}=R\times {\mathcal B(\eta_c \rightarrow X)\over \mathcal B(J/\psi \rightarrow \mu^+\mu^-)},
\end{eqnarray}
where $\mathcal B(\eta_c \rightarrow X)$ is the branching fraction of the $\eta_c$ in an exclusive decay channel  $X$. $R$ is the efficiency-corrected yield ratio, 
\begin{eqnarray}
R={N_{J/\psi}\over N_{\eta_c}}\times {\epsilon_{\eta_c}\over \epsilon_{J/\psi}},
\end{eqnarray}
where $N_{J/\psi}$ and $N_{\eta_c}$  are the observed $J/\psi$ and $\eta_c$ yields. $\epsilon_{\eta_c}$ and $ \epsilon_{J/\psi}$ the total efficiencies, which are determined by the combination of the simulated and calibration samples. The value of  $R$ should be obtained from $ \Lambda_b\rightarrow J/\psi p K^-$ and $\Lambda_b\rightarrow \eta_c p K^-$ simulated samples, both selected using the same criteria used in data \cite{Aaij:2018bla}. So far, we have no information about the above parameter. However, the LHCb collaboration has successfully triggered  the $\eta_c$ events  \cite{Aaij:2018bla}. In 2018, the decay process $B^0\rightarrow\eta_cK^+\pi^-$ was observed by the LHCb collaboration \cite{Aaij:2018bla}. It enhanced our confidence in searching the $P_c$ states in the $\eta_cp$ decay mode.  At last, the LHCb collaboration will collect large data  in the Upgrade II era, which may help the search of the states decaying into the final states with the $\eta_c$ \cite{Bediaga:2018lhg}.\\

In contrast to the prediction in scenario I, the $P_c(4440)$ may be a little hard to be observed in the $\eta_c p$ channel in scenario II due to the destructive interference of the configurations $\Sigma_c \bar D^*$ and $\Sigma^*_c \bar D^*$. The tensions between the two  scenarios will help to probe the molecular components in the $P_c(4312)$ and $P_c(4440)$ states.

In Table \ref{results}, we find that the predicted partial decay widths in scenario II are generally larger than those in scenario I except  that of $P_c(4440)\rightarrow \eta_c p$. This may come from the different relative molecular wave functions in the two scenarios. In scenario I, the radial wave function in the momentum space $|\mathbf p| R_{S0}(|\mathbf p|) $ ($R_{S0}( |\mathbf p|)$ is the molecular wave function in Eq. (\ref{rel}) with $ \mathbf p$ being the relative momentum) is zero at $|\mathbf p|=0$, while in scenario II the wave function obtained in the OBE model  is not zero after Fourier transformation of the wave functions in Fig.~\ref{fig:pc}. In the overlap of the wave functions, the region with small $|\mathbf p|$ may play an important role.

The obtained decay widths and branching fraction ratios are useful to explore the molecular assignment for the hidden-charm pentaquark states, which can be also examined by the experiments and the lattice QCD in the coming years.  Very recently, the LHCb collaboration searched the $\eta_c p$ mode in the
$\Lambda_b \to \eta_c p K^-$ decay with no significant signal of the $P_c(4312)$ resonance \cite{Aaij:2020mlx}.~It is interesting to investigate whether the couple-channel effect distorts the predicted decay pattern. The investigation of the decay properties will help to explore the reasonability of the molecular assignment  and understand the inner dynamics of the exotic states.

\section*{Acknowledgements}

The authors are very grateful to X. L. Chen and W. Z. Deng for very helpful discussions. This project is supported by the National Natural Science Foundation of China under Grants 11575008, 11621131001, 11975033 and 11975165. G. J. Wang is also supported by the China Postdoctoral Science foundation No. 2019M660279. R. C. is also supported by the National Postdoctoral Program for Innovative Talent. This work is partly supported by the China National Funds for Distinguished Young Scientists under Grant No. 11825503 and the National Program for Support of Top-notch Young Professionals.

\begin{appendix}
\section{The spatial wave function $\phi$ }\label{Appendixwf}

In this work, we use the Gaussian basis to mimic the spatial wave functions of the hadrons. The wave function of an S-wave meson in the momentum space reads
\begin{eqnarray} \label{GW}
\phi(\mathbf p_{\text{rel}})=\frac{1}{\beta^{3/2}\pi^{3/4}}\text{exp}\left(-\frac{\mathbf p_{\text{rel}}^{2}}{2\beta^{2}}\right),
\end{eqnarray}
with the reduced momentum
\begin{eqnarray}
\mathbf p_{\text {rel}}=\frac{m_{\bar{q}}\mathbf{p}_{q}-m_{q}\mathbf{p}_{\bar{q}}}{m_{q}+m_{\bar{q}}},
\end{eqnarray}
where $m_q(m_{\bar{q}})$ and $\mathbf{p}_q$ $(\mathbf{p}_{\bar{q}})$ are the mass and momentum of the quark (antiquark) in the meson, respectively. $\beta$ is the oscillating parameter. Its value is estimated with the root mean square radius of the meson in the Godfrey-Isgur model \cite{Godfrey:1985xj}, which are listed in Table \ref{par:wave}.

\renewcommand\tabcolsep{0.4cm}
\renewcommand{\arraystretch}{1.8}
\begin{table}[!htbp]
\caption{The oscillating parameters (in units of GeV) in the wave functions of the mesons and baryons. The superscripts $c$ and $b$ represent the charmed and bottom baryons, respectively. }\label{par:wave}
\begin{tabular}{cccccc}
\toprule[1pt]
$\alpha_{\rho}$ & $\alpha_{\lambda}^{c}$ & $\alpha_{\lambda}^{b}$ & $\beta_{D}$ & $\beta_{D^{*}}$ & $\beta_{\eta_{c}}$ \\
\hline
$0.40$ & $0.49$ & $0.51$ & $0.60$ & $0.52$ & $0.83$ \\
\bottomrule[1pt]
$\beta_{J/\psi}$ & $\beta_{B}$ & $\beta_{B^{*}}$ & $\beta_{\eta_{b}}$ & $\beta_{\Upsilon(1S)}$\\
\hline
$0.73$ & $0.58$ & $0.54$ & $1.22$ & $1.14$\\
\bottomrule[1pt]
 \end{tabular}
\end{table}

For an S-wave baryon with two independent Jacobi coordinates, the wave function is
\begin{eqnarray}
\phi(\mathbf {p_3},\mathbf {p_4},\mathbf {p_5})=\frac{3^{3/4}}{\pi^{3/2}\alpha_{\rho}^{3/2}\alpha_{\lambda}^{3/2}}\exp\left(-\frac{\mathbf p_{\rho}^{2}}{2\alpha_{\rho}^{2}}-\frac{\mathbf  p_{\lambda}^{2}}{2\alpha_{\lambda}^{2}}\right),\quad
\end{eqnarray}
with
\begin{eqnarray}
&&\mathbf p_{\rho}=\frac{\left(m_{3}+2m_{5}\right)\mathbf{p_{4}}-\left(m_{3}+2m_{4}\right)
\mathbf{p_{5}}+\left(m_{5}-m_{4}\right)\mathbf{p_{3}}}{\sqrt{2}(m_{4}+m_{5}+m_{3})}, \nonumber \\
&&\mathbf p_{\lambda}=\sqrt{\frac{3}{2}}\frac{m_{3}\left(\mathbf{p_{4}}+\mathbf{p_{5}}\right)-\left(m_{4}+m_{5}\right)
\mathbf{p_{3}}}{m_{4}+m_{5}+m_{3}},
\end{eqnarray}
where $m_3$, $m_4$, and $m_5$ are the masses of the consistent quarks. $\alpha_{\rho}$ and $\alpha_{\lambda}$ are the oscillating parameters, which satisfy
\begin{eqnarray}
\label{lambda}
\alpha^2_{\lambda}=\sqrt{\frac{3m_Q}{2m_q+m_Q}}\alpha^2_{\rho},
\end{eqnarray}
for a heavy baryon. Here, $m_q$ and $m_Q$ are the masses of the light and heavy quarks, respectively.
For the nucleon with three light quarks, one has $\alpha_{\rho}=\alpha_{\lambda}=\alpha$ and its wave function can be written as
\begin{eqnarray}
\phi(\mathbf {p_3},\mathbf {p_4},\mathbf {p_5})=\frac{3^{3/4}}{\pi^{3/2}\alpha^{3}}\exp\left(-\frac{\mathbf p_{\rho}^{2}+\mathbf p_{\lambda}^{2}}{2\alpha^{2}}\right).
\end{eqnarray}
The parameter $\alpha_{\rho}$ in the wave functions for the $du/uu$ systems is taken as $\alpha_{\rho}=0.4$ GeV. The values of the  $\alpha_{\lambda}$ are derived using Eq. (A5) and the results are collected in Table \ref{par:wave}.

\section{The flavor-spin factor $\mathcal I_{\text{flavor-spin}}$}\label{Appendixsf}

In the quark interchange model, the flavor-spin factor $\mathcal I_{\text{flavor-spin}}$ reads
\begin{widetext}
\begin{eqnarray}
\mathcal I_\text{flavor-spin}=\left\langle\left[\chi_{C}(13)_{s_{c}}^{I_{c}}\otimes\chi_{D}(245)_{s_{d}}^{I_{d}}
\right]_{S^{'}}^{I^{'}}|\hat V_s|\left[\chi_{A}(12)_{s_{a}}^{I_{a}}\otimes
\chi_{B}(345)_{s_{b}}^{I_{b}}\right]_{S}^{I}\right\rangle,
\end{eqnarray}
\end{widetext}
where $S^{(')}$ and $I^{(')}$ are the spin and isospin of the initial (final) state, respectively. $s_{a,b,c,d}$ and $I_{a,b,c,d}$ denote the spin and isospin for the different hadrons. $\hat V_s$ represents the spin operator. Here, $\chi$ contains the wave functions in the spin and flavor space. 

We use the scattering process $\bar D^{-(*)}\Sigma_{c}^{(*)++}\rightarrow J/\psi(\eta_{c})p$ as an example to illustrate the calculation of the factor $\mathcal I_{\text{flavor-spin}}$. In the scattering process, the flavor factor $\mathcal I_{\text{flavor}}=1$.
For the spin factor $\mathcal I_{\text{spin}}$, when $\hat V_s=\mathbf{1}$, it reads
\begin{widetext}
\begin{eqnarray}
\mathcal I_{\text{spin}}=\langle\left[\chi_{C}(13)_{s_{c}}\otimes\chi_{D}(245)_{s_{d}}\right]_{S^{'}}
|\mathbf{1}|\left[\chi_{A}(12)_{s_{a}}\otimes\chi_{B}(345)_{s_{b}}\right]_{S}\rangle =\delta_{SS'}\hat{s_{c}}\hat{s_{d}}\hat{s_{a}}\hat{s_{b}}\left(\begin{array}{ccc}
\frac{1}{2} & \frac{1}{2} & s_{c}\\
\frac{1}{2} & s_{45} & s_{d}\\
s_{a} & s_{b} & S
\end{array}\right),
\end{eqnarray}
with $\hat X=\sqrt{2X+1}$. $s_{45}$ is the spin of the fourth and fifth quarks.  The matrix in the right-hand side represents the Wigner 9j-symbol. When $\hat V_s=\mathbf{S}_{i}\cdot\mathbf{S}_{j}$, the $\mathcal I_{\text{spin}}$ for different diagrams in Fig. \ref{fig:qfill} are expressed as
\begin{eqnarray*}
 \mathcal  I^{d_1}_{\text{spin}} &=& \langle\left[\chi_{C}(13)_{s_{c}}^{I_{c}}\otimes\chi_{D}(245)_{s_{d}}^{I_{d}}
 \right]_{S^{'}}^{I^{'}}|\mathbf{\mathbf{S}_{2}\cdot\mathbf{S}_{3}}|
 \left[\chi_{A}(12)_{s_{a}}^{I_{a}}
 \otimes\chi_{B}(345)_{s_{b}}^{I_{b}}\right]_{S}^{I}\rangle\nonumber\\
  &=& \sum_{s_{23}s_{145}}\delta_{SS'}(-1)^{s_{c}+s_{a}}\hat{s}_{a}\hat{s}_{b}\hat{s}_{c}\hat{s}_{d}
  \hat{(s_{23}}\hat{s_{145}})^{2}\left(\begin{array}{ccc}
\frac{1}{2} & \frac{1}{2} & s_{c}\\
\frac{1}{2} & s_{45} & s_{d}\\
s_{32} & s_{145} & S
\end{array}\right)\left(\begin{array}{ccc}
\frac{1}{2} & \frac{1}{2} & s_{a}\\
\frac{1}{2} & s_{45} & s_{b}\\
s_{23} & s_{145} & S
\end{array}\right)(-1)^{s_{23}-1}(\frac{1}{2}s_{23}(s_{23}+1)-\frac{3}{4}),\\
\mathcal I^{d_2}_{\text{spin}} &=& \langle\left[\chi_{C}(13)_{s_{c}}^{I_{c}}\otimes\chi_{D}(245)_{s_{d}}^{I_{d}}
 \right]_{S^{'}}^{I^{'}}|\mathbf{\mathbf{S}_{2}\cdot\mathbf{S}_{4}}|
 \left[\chi_{A}(12)_{s_{a}}^{I_{a}}\otimes\chi_{B}(345)_{s_{b}}^{I_{b}}\right]_{S}^{I}\rangle\nonumber\\
 &=& \sum_{s_{24}}\delta_{SS'}(-1)^{3+2s_{d}}\hat{s_{c}}\hat{s_{d}}\hat{s_{a}}\hat{s_{b}}\left(\begin{array}{ccc}
\frac{1}{2} & \frac{1}{2} & s_{a}\\
\frac{1}{2} & s_{45} & s_{b}\\
s_{c} & s_{d} & S
\end{array}\right)(\hat{s_{24}}\hat{s_{45}}\left\{ \begin{array}{ccc}
1/2 & 1/2 & s_{24}\\
1/2 & s_{d} & s_{45}
\end{array}\right\} )^{2}(\frac{1}{2}s_{24}(s_{24}+1)-\frac{3}{4}),\\
\mathcal I^{\bar{d_{1}}}_{\text{spin}} &=& \langle\left[\chi_{C}(13)_{s_{c}}^{I_{c}}\otimes\chi_{D}(245)_{s_{d}}^{I_{d}}
 \right]_{S^{'}}^{I^{'}}|\mathbf{S}_{1}\cdot\mathbf{S}_{3}|
 \left[\chi_{A}(12)_{s_{a}}^{I_{a}}\otimes\chi_{B}(345)_{s_{b}}^{I_{b}}\right]_{S}^{I}\rangle\nonumber\\
 &=&\delta_{SS'} \hat{s_{c}}\hat{s_{d}}\hat{s_{a}}\hat{s_{b}}\left(\begin{array}{ccc}
\frac{1}{2} & \frac{1}{2} & s_{c}\\
\frac{1}{2} & s_{45} & s_{d}\\
s_{a} & s_{b} & S
\end{array}\right)(\frac{1}{2}s_{c}(s_{c}+1)-\frac{3}{4}),\\
\mathcal I^{\bar{d_{2}}}_{\text{spin}} &=& \langle\left[\chi_{C}(13)_{s_{c}}^{I_{c}}\otimes
 \chi_{D}(245)_{s_{d}}^{I_{d}}\right]_{S^{'}}^{I^{'}}|\mathbf{\mathbf{S}_{1}\cdot
 \mathbf{S}_{4}}|\left[\chi_{A}(12)_{s_{a}}^{I_{a}}\otimes\chi_{B}(345)_{s_{b}}^{I_{b}}
 \right]_{S}^{I}\rangle\nonumber\\
  &=& \sum_{s_{23}s{}_{145}}\sum_{s_{14}}\delta_{SS'}(-1)^{s_{c}+s_{a}}\hat{s}_{a}\hat{s}_{b}
  \hat{s}_{c}\hat{s}_{d}\hat{(s_{23}}\hat{s_{145}})^{2}\left(\begin{array}{ccc}
\frac{1}{2} & \frac{1}{2} & s_{c}\\
\frac{1}{2} & s_{45} & s_{d}\\
s_{32} & s_{145} & S
\end{array}\right)\left(\begin{array}{ccc}
\frac{1}{2} & \frac{1}{2} & s_{a}\\
\frac{1}{2} & s_{45} & s_{b}\\
s_{23} & s_{145} & S
\end{array}\right)\nonumber\\
 &&\times(-1)^{s_{23}-1}\left((-1)^{3/2+s_{145}}\hat{s_{14}}\hat{s_{45}}\left\{ \begin{array}{ccc}
1/2 & 1/2 & s_{14}\\
1/2 & s_{145} & s_{45}
\end{array}\right\} \right)^{2}(\frac{1}{2}s_{14}(s_{14}+1)-\frac{3}{4}),
\end{eqnarray*}
\end{widetext}
where the symbol $\Big\{ \Big\}$ represents the Wigner's 6j-symbol. With the method, one can easily calculate $\mathcal I_{\text{spin}}$ for the post diagrams.

\section{The decay of the $P_b$ state }\label{Appendixpb}
\begin{figure*}[htbp]
\centering
\subfigure[~$P_b(\Sigma_b B)$]{
\begin{minipage}[t]{0.33\linewidth}
\centering
\includegraphics[width=1\textwidth]{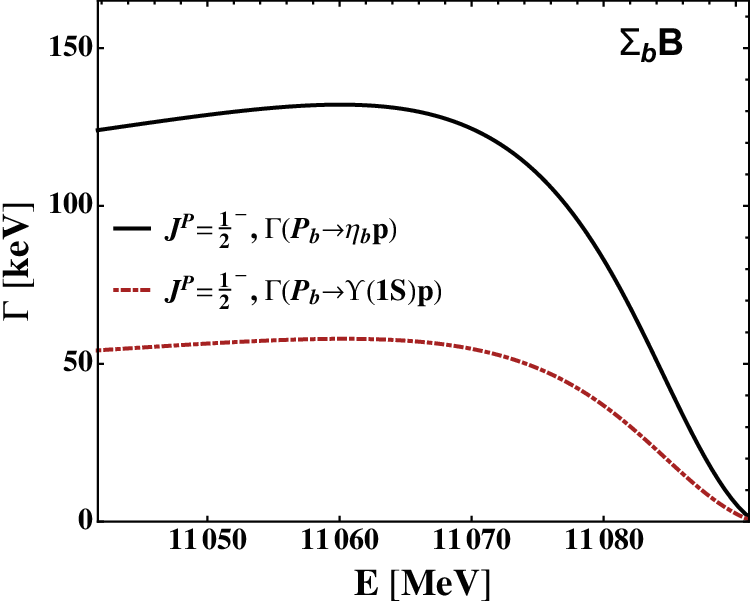}

\end{minipage}%
}%
\subfigure[~$P_b(\Sigma_bB^*)$]{
\begin{minipage}[t]{0.33\linewidth}
\centering
\includegraphics[width=1\textwidth]{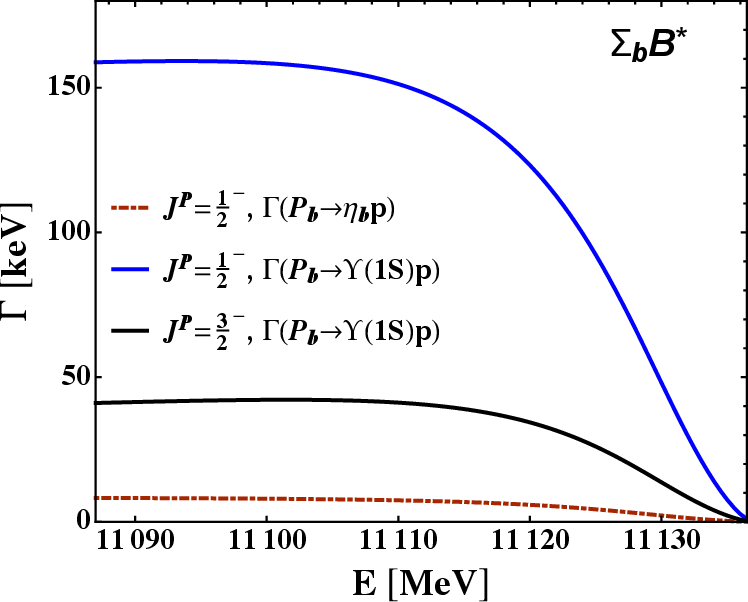}

\end{minipage}%
}%
\\
\subfigure[~$P_b( \Sigma^*_bB)$ ]{
\begin{minipage}[t]{0.335\linewidth}
\centering
\includegraphics[width=1\textwidth]{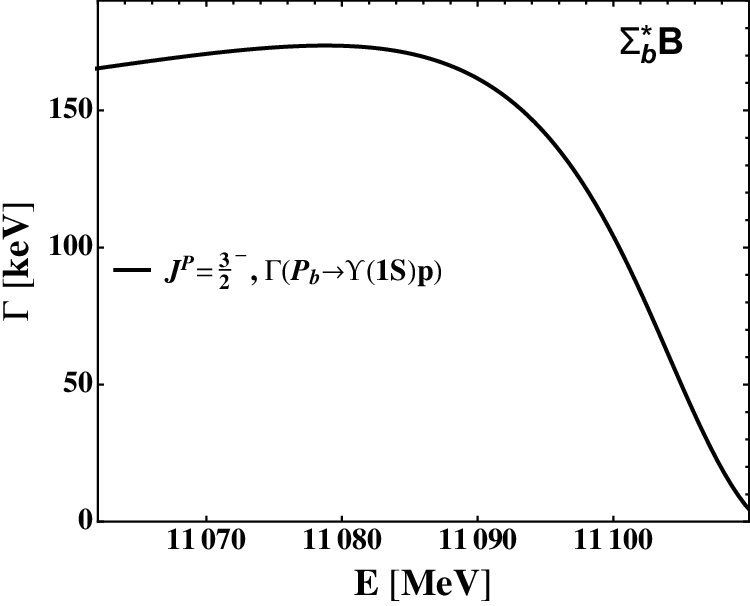}
\end{minipage}
}%
\subfigure[~$P_b(\Sigma^*_bB^*)$ ]{
\begin{minipage}[t]{0.33\linewidth}
\centering
\includegraphics[width=1\textwidth]{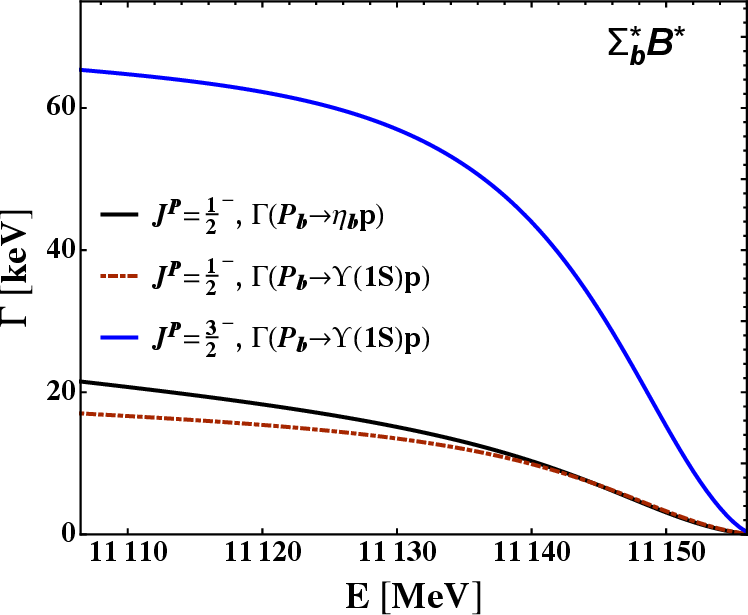}
\end{minipage}
}%
\centering
\caption{ The mass dependence of the partial decay widths  of the $\Sigma^{(*)}_bB^{(*)}$ molecular states if they exit. The molecular wave function is estimated by a Gaussian function. The corresponding binding energy is in the range of $ -50\sim -1$ MeV. }
\label{fig:massdenpendenceB}
\end{figure*}

\begin{figure*}[!htp]
\centering
\includegraphics[width=1\textwidth]{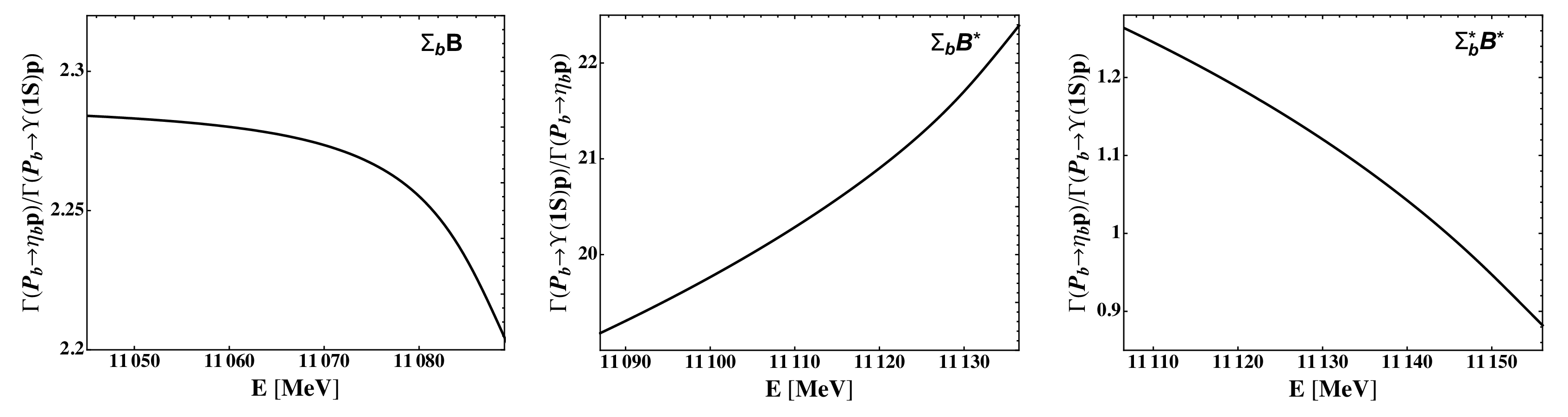}
\caption{ The branching fraction ratios for the $P_b$ states with $J^P=\frac{1}{2}^-$ decaying into the $\Upsilon(1S)p$ and $\eta_b p$ channels. }
\label{fig:ratioB}
\end{figure*}

In this section, we assume that the pentaquark states in the bottom sector exist as the pure molecules $B^{(*)}\Sigma^{(*)}_b$.~We extend the method in scenario I to calculate the partial decay widths from the $P_b$ states into the $\Upsilon(1S) p$ and $\eta_bp$ channels. The decay widths and the decay ratios between the two channels versus the mass spectra of the $P_b$ states are shown in Figs. \ref{fig:massdenpendenceB} and \ref{fig:ratioB}, respectively. The results in Figs.~\ref{fig:massdenpendenceB} and \ref{fig:ratioB} show that the partial decay width is sensitive to the initial masses of the pentaquark states, which has not been observed in the experiments. So far, we cannot give more discussion about their decay widths. However, decay ratios are relatively stable and not sensitive to the initial masses of the pentaquark states.

In scenario II, we also investigate the $\Sigma^{(*)}_b B^{(*)}$ system using the one-boson-exchange model. However, the existence and the inner structures of the $P_b$ states are very sensitive to the cutoff parameter $\Lambda$, which may induce quite different decay patterns. Thus, we do not perform the numerical calculation and cannot explore the influences of the coupled channel effect without experimental data.

Nevertheless, we hope our results in scenario I may serve as a hint for the search of the $P_b$ states in the future experiments. 
\\

\end{appendix}

\end{document}